\documentclass[nolinenumbers, twocolumn, trackchanges]{aastex631}
\usepackage{textgreek, graphicx}
\usepackage{amsmath}
\usepackage[inkscapelatex=false]{svg}
\usepackage{threeparttable}
\usepackage{booktabs}
\usepackage{physunits}
\usepackage{wrapfig}

\newcommand\omicron{o}

\begin{document}

\title{Evolution of the disk in the Be binary $\delta$ Scorpii probed during three periastron passages}

\author[0009-0007-9595-2133]{R.G. Rast}
\affiliation{Department of Physics and Astronomy, The University of Western Ontario, London, ON N6A 3K7, Canada}

\author[0000-0001-9900-1000]{C.E. Jones}
\affiliation{Department of Physics and Astronomy, The University of Western Ontario, London, ON N6A 3K7, Canada}

\author[0000-0002-9369-574X]{A.C. Carciofi}
\affiliation{Instituto de Astronomia, Geof\'ica e Ci\'encias Atmosf\'ericas, Universidade de S\~ao Paulo, Brazil}

\author[0000-0003-0696-2983]{M.W. Suffak}
\affiliation{Department of Physics and Astronomy, The University of Western Ontario, London, ON N6A 3K7, Canada}

\author[0000-0002-6073-3906]{A.C.F. Silva}
\affiliation{Instituto de Astronomia, Geof\'ica e Ci\'encias Atmosf\'ericas, Universidade de S\~ao Paulo, Brazil}

\author[0000-0003-4155-8513]{G.W. Henry}
\affiliation{Center of Excellence in Information Systems, Tennessee State University, Nashville, TN 37209, USA}

\author{C. Tycner}
\affiliation{Department of Physics, Central Michigan University, Mt. Pleasant, MI 48859, USA}



\begin{abstract}

We examine the evolution of the disk surrounding the Be star in the highly eccentric binary system $\delta$ Scorpii over its three most recent periastron passages. $V$-band and $B-V$ photometry, along with H$\alpha$ spectroscopy are combined with a new set of extensive multi-band polarimetry data to produce a detailed comparison of the disk’s physical conditions during the time periods surrounding each closest approach of the secondary star. We use the three-dimensional Monte Carlo radiative transfer code \textsc{HDUST} and smoothed particle hydrodynamics (\textsc{SPH}) code to support our observations with models of disk evolution, discussing the behaviour of the H$\alpha$ and He\,\textsc{i} 6678 lines, $V$-band magnitude, and polarization degree. We compare the characteristics of the disk immediately before each periastron passage to create a baseline for the unperturbed disk. We find that the extent of the H$\alpha$ emitting region increased between each periastron passage, and that transient asymmetries in the disk become more pronounced with each successive encounter. Asymmetries of the H$\alpha$ and He\,\textsc{i} 6678 lines in 2011 indicate that perturbations propagate inward through the disk near periastron. When the disk's direction of orbit is opposite to that of the secondary, the parameters used in our models do not produce spiral density enhancements in the H$\alpha$ emitting region because the tidal interaction time is short due to the relative velocities of the disk particles with the secondary. The effects of the secondary star on the disk are short-lived and the disk shows independent evolution between each periastron event.

\end{abstract}



\section{Introduction} \label{sec:intro}

Be stars are non-supergiant stars of spectral class B that have exhibited Balmer lines in emission at some point in their lifetime \citep{jas81, col87}. Classical Be stars are more narrowly defined as main sequence (or slightly evolved) B-type stars whose Balmer emission lines originate from circumstellar disks of gas that are rotating in Keplerian fashion and fed by the central star \citep{bjo12}. These stars rotate at rates greater than any other non-degenerate stars \citep{tow04}. Their rotation rates are usually thought to be more than 75\% of the critical velocity, where the gravitational force is balanced by the centrifugal force at the equator \citep{riv13, tow04, fre05}. In addition, the stars exhibit multi-mode, non-radial pulsations that may play a role in the star-disk mass transfer \citep{baa18, baa20, lab22}.

The disks surrounding classical Be stars have been successfully described using the viscous decretion disk model \citep{lee91}. Dynamically, this model is similar to the one for viscous accretion disks, only instead of gas flowing inward onto the star as is the case with accretion, the gas flows outward. The central star contributes mass and angular momentum via parcels of gas to the inner disk. The gas is then launched into Keplerian orbit and drifts outward facilitated by viscosity. This model has been used in the radiative transfer code \textsc{HDUST} to predict observables of Be stars at specific times \citep{car06a}. For example, \citet{dea20} used \textsc{HDUST} to successfully predict the spectral energy distribution (SED), $V$-band polarization angle, as well as the H$\alpha$ and Br$\gamma$ line profiles of the star $\omicron$ Aqr. Similarly, \textsc{HDUST} was used by \citet{gho21} to reproduce observational trends in photometry, spectra and broad-band polarimetry for $\omega$ CMa.

Among Be stars, the H$\alpha$ line (6562 \r{A}) may be singly-peaked (usually observed more pole-on) or doubly-peaked (at higher inclinations when viewing Doppler shifted material) \citep{han89, hum94, sig20, sig23}. Edge-on stars can also exhibit shell characteristics, with deep absorption cores that may be flanked by emission peaks. Similar to the other H\,\textsc{i} lines, the H$\alpha$ emission line is formed predominantly through recombination in the disk \citep{por03, bjo12}. For an optically thick disk, the strength of the H$\alpha$ flux is directly proportional to the area of the emitting region \citep{tyc05}, which constitutes a lower limit on the extent of the circumstellar disk. Information about the inner regions of the disk may be carried in the He\,\textsc{i} 6678 \r{A} line, which is produced near the Be star.

In many Be stars, the equivalent width (EW) and full width at half maximum (FWHM) of the H$\alpha$ emission line have been observed to vary over time, indicating changes in the size  of the H$\alpha$ emitting region or changes in the density of the emitting volume. 
The strength of the H$\alpha$ line can be used to track the evolution of a star as it acquires a circumstellar disk and enters its Be phase, or as it loses the disk to appear as a typical B star \citep{wis10}. For example, $\omega$ Ori was observed to build a disk over a period of a few years \citep{gui84}. By contrast, the disks of 60 Cyg and \textpi{} Aqr dissipated over periods of roughly 1000 days and 2440 days, respectively \citep{wis10}, while the disk of 66 Oph has been dissipating for more than 20 years \citep{mar21}. Over a few decades, the Be star $\gamma$ Cas famously underwent two shell phases \citep{bal40, bor20} where the disk may have been dynamically destroyed \citep{baa23} before regaining the disk and remaining relatively stable in recent years \citep{bor20}.

The shape of the H$\alpha$ profile also encodes information about the density distribution (and velocites) of the disk. The ratio of the relative flux of the violet and red peaks (V/R) of double-peaked lines may vary cyclically over 5 to 10 years \citep{oka97, ste09}. The cycle lengths and amplitudes of the variations may also change over time \citep{ruz09}. The most successful and widely accepted model to address this involves one-armed spiral density waves in the inner disk \citep{oka97, car09}. Density enhancements have been directly observed \citep{vak98}, supporting this model. 

Photometric observations of Be stars can also reveal important information about disk size and mass-loss rates. $V$-band photometry is of particular interest in Be stars. The majority of the $V$-band flux originates in the inner regions of the disk, typically within two stellar radii ($\mathrm{R_{\star}}$) with the exact emitting regions depending on disk density \citep{car11, riv13}. Therefore, the observations in the $V$-band are the fastest indicators to signal changes in the disk's size or density due to photospheric activity such as outbursts \citep{car11}. For example, \citet{wis10} found that changes in the  H$\alpha$ profile, which forms in a more extended volume of the disk, lagged behind the $V$ band by hundreds of days. 

Together with the polarization angle, changes in the polarization degree have been attributed to mass ejections and resulting density alterations of the disk \citep{car07}. The polarization signature originates in the disk, where unpolarized light undergoes Thompson scattering \citep{woo97}. The relative strength of the polarization, polarization position angle, and wavelength-dependence of the polarization give information about the disk geometry, number of scatterers in the disk, and the density of the disk, respectively \citep{hau14}. During disk growth, the polarization in the Balmer jump rises rapidly while the $V$-band polarization degree increases at a slower rate \citep{hal13}. Conversely, during dissipation events, the polarization in the Balmer jump drops rapidly while the $V$-band polarization degree takes longer to decrease \citep{hal13}. Since the Balmer jump forms at smaller disk radii than the $V$-band, this suggests disk growth is caused by mass loss near the stellar surface, while the inner regions of the disk are the first to disappear during dissipation events. This pattern has been observed in $\omega$ Ori \citep{son88}.

Be stars, like other high-mass stars, are commonly found in binary systems. In a study of 340 Be stars, \citet{mir16} concluded that more than 50\% of Be stars brighter than $V$ $\sim$ 4 mag were in binary systems. \citet{kle19} studied the infrared and radio emissions from 57 classical Be stars and found indirect evidence of binarity in 26 of these objects. \citet{dod24} found binary fractions of 28\% for B and 17\% for Be stars at separations between 0.02 and 0.2 arcsec. They attribute the relative lack of binary Be stars at these separations to the presence of stripped companions, which constitute an increasing fraction of detected secondary stars in Be binary systems \citep{mar23} and would not have been detected by their methods \citep{dod24}. 

Tidal interactions between the secondary star and the disk may result in structural distortions or truncations of the disk. Two-armed waves have been proposed in connection to such tidal perturbations \citep{oka02}. They are associated with much smaller V/R amplitudes that are phase-locked with the binary orbital period \citep{pan18}. For example, the H$\alpha$ profile of $\pi$ Aqr has shown radial velocity and V/R variations that are phase-locked to the orbital period of the system \citep{bjo02, zha13}. \citet{mir23} suggest that these V/R variations can be used to detect binarity in Be stars if other methods are inconclusive. \citet{oka02} studied a subgroup of Be/X-ray systems where the disk was aligned to the orbital plane, and found that the density within the truncation radius (which is located where the tidal torque and viscous torque are balanced) was larger than in the untruncated disks of single Be stars. In eccentric orbits, the truncation radius is smaller at periastron, and a tightly-wound spiral structure may be generated at closest approach \citep{oka02}. \citet{cyr20} showed that for large misalignment angles and higher viscosities, the spiral arms are folded more closely. If the eccentricity of the system is close to one or if the disk and orbital plane are misaligned by more than 60$^\circ$, these truncation effects are inefficient \citep{mar11, oka07}.

$\delta$ Scorpii ($\delta$ Sco) is a bright spectroscopic binary system consisting of a 15 M$_\odot$ Be star and an 9 M$_\odot$ secondary. While the presence of the secondary star has been confirmed through interferometry, its spectral lines have yet to be directly detected as its brightness is roughly one-fifth that of the primary \citep{bed93, tan09, hut21}. As a result, its spectral class has not been determined to a high degree of certainty \citep{bod20}, although observations are compatible with a main sequence star \citep{tan09, car06b, mir13}. This fact makes $\delta$ Sco of special interest, as a literature study by \citet{bod20} found a lack of verified main sequence companions among Be stars earlier than B1.5. The inclinations of the primary star and its disk have been measured through interferometry to be approximately 26$^{\circ}$ \citep{mei13} and 27$^{\circ}$ \citep{che12}, respectively. Similarly, \citet{car06b} found a stellar inclination of 38$^{\circ}$ $\pm$ 5$^{\circ}$ using polarimetry models that assumed the disk was equatorial.

This system is highly eccentric (e $\approx$ 0.94) with a period of approximately 10.8 years \citep{tan09, tyc11, mei11}. At periastron, \citet{tyc11} have found the minimum separation between the two stars to be 6.14 mas. Adopting a distance of 150 pc, as determined by astrometric parallax \citep{van07}, this corresponds to a separation of 0.92 AU or 29 $R_{\star}$, where $R_{\star}$ represents the radius of the primary star.

The primary star in this system represents one of few examples where a Be star has been consistently observed as its disk grew and evolved over three decades. \textdelta{} Sco was first reported to display faint H$\alpha$ emission in 1990 \citep{cot93} and showed strong  H$\alpha$ emission by the summer of 2000 \citep{mir01}. Since then, the binary system has undergone three periastron passages: on UT 2000 September 10 \citep{tyc11}, UT 2011 July 3 \citep{che12}, and UT 2022 April 24 (based on the orbital period of \citealt{tyc11}). The behaviour of the system during its periastron passages in 2000 and 2011 has been studied by \citet{mir01}, \citet{tyc11}, \citet{mir13}, \citet{che12} and others. \citet{suf20} studied long-term varations of the system over nearly two decades from 2000 to 2018, finding significant disk growth during 2009 to 2011 as well as evidence of partial disk dissipation at apastron. The behaviour of the system during its most recent periastron in April of 2022 has yet to be studied.

In this paper, we present for the first time a systematic comparison of the behaviour of the disk surrounding the Be component of $\delta$ Sco during the binary system's three most recent periastra. We discuss how the evolution of the disk between each periastron passage influences the interactions between the secondary star and the disk at closest approach. We utilize H$\alpha$ spectroscopy, $V$-band photometry, and multiband polarimetry to probe disk structure, size and density, supporting these observations with 3-dimensional simulations and radiative transfer calculations. Section \ref{sec:methods} details our methodology, including a description of the codes used in our analysis. Section \ref{sec:obs} provides our observations and their sources, along with any reduction methods applied. Our findings are presented in Section \ref{sec:results}, while Section \ref{sec:discussion} contains our discussion and  Section \ref{sec:conclusion} provides our conclusions.

\section{Methodology} \label{sec:methods}

Our model predictions were accomplished using two three-dimensional codes: \textsc{HDUST}, used to investigate the behaviour of the $V$-band polarimetry in the 2010s, and a smoothed particle hydrodynamics (\textsc{SPH}) code, used to trace the structure and extent of the disk through the three close encounters between the secondary and primary.

\textsc{HDUST} is a non-local thermodynamic equilibrium (NLTE) radiative transfer code that uses the Monte Carlo method to predict observables for the models of circumstellar disks \citep{car06a}. Given particular input parameters of a Be star-disk system, such as stellar mass, radius and rotation rate, disk radius, and disk density, \textsc{HDUST} simulates the random propagation of photon packets through the disk medium. It calculates the temperature structure of the disk, along with the ionization fraction and level populations in the NLTE regime. \textsc{HDUST} can predict images, spectral lines, photometry, and polarization in wavelengths of interest. It has been used for studies on individual Be stars (for example, see \citealt{suf20}, \citealt{ric21}, \citealt{gho21}, \citealt{mar21}, \citealt{mar22}, \citealt{rub23}) and for large statistical studies (for example, see \citealt{rim18} and \citealt{suf23}).  

In our \textsc{HDUST} models, we used a power law for volume density of the disk 
\begin{equation}
    \rho(r,z) = \rho_{0} \left(\frac{R_{\star}}{r}\right)^n \exp \left(- \frac{z^2}{2H^2}\right)\,,
    \label{eq:power_law}
\end{equation}
where $r$ is the radial position in the disk, $z$ is the vertical position in the disk, $R_{\star}$ is the equatorial radius of the Be star, $\rho_0$ is the density where $r$ = $R_{\star}$ and $z = 0$, $n$ is the parameter that governs how quickly the density drops off with increasing $r$, and $H$ represents the scale height.

The \textsc{SPH} code was first developed by \citet{ben90a} and \citet{ben90b}. \citet{bat95} modified the code to increase efficiency by adding a Runge-Kutta integrator, and \citet{oka02} further adapted it for the viscous decretion disks of Be binary systems. It has been used extensively to follow Be star disk structure over time. For recent work, see \citet{pan16}, \citet{cyr17} and \citet{suf22}.

The \textsc{SPH} simulations begin with a diskless system, given the values for the mass and radii of both stellar components and the orbital parameters of the system. We then inject material into the disk at a prescribed rate near the primary star. The disk material flows slowly outward facilitated by viscosity following the prescription of \citet{sha73}:

\begin{equation}
    \nu = \alpha c_s H\,,
\end{equation}

\noindent where $\nu$ is the viscosity, $\alpha$ is a dimensionless parameter that scales the viscosity (hereafter called the viscosity parameter), $c_s$ is the sound speed in the disk, and $H$ is the scale height.  

\section{Observations and Data} \label{sec:obs}

\subsection{Photometry}

Photometric data for 2009 to 2023 were collected at the Fairborn Observatory in the Patagonia Mountains of southern Arizona, USA. These observations were acquired using the T3 0.4\,m automatic photoelectric telescope (APT) equipped with a photometer that measures photometric count rates with an EMI 9924B photomultiplier tube. These observations each have an accuracy of approximately 0.003 to 0.005 mag on a good night, with seasonal brightness means of 0.0001 to 0.0002 mag \citep{jon11, hen99}. Observations were recorded in the Johnson $B$ and $V$ bandpasses. The data spanning 2009 to 2019 were first published in \citet{suf20}. 

Additional data for 2000 to 2023 were obtained from the American Association of Variable Star Observers (AAVSO)\footnote{\href{https://www.aavso.org/data-access}{https://www.aavso.org/data-access}}. To ensure high-quality measurements, we selected $V$-band values that were flagged as ``validated," meaning they were peer reviewed by AAVSO members. Together with the observations from Fairborn, these data created excellent coverage for the periastron years of interest as well as the time period between.

\subsection{Spectroscopy}

Spectral data for 2000 were obtained from the Ritter Observatory Public Archive. They were observed at the Ritter Observatory in Toledo, Ohio, USA, using a 1.06\,m telescope equipped with a fiber-fed echelle spectrograph and a CCD camera manufactured by Wright Instruments Ltd.  These spectra have a resolving power of $R$ = 26,000 over a wavelength range of 5285 to 6527 \r{A}, split across 9 orders. We focused on the order containing the H$\alpha$ line, ranging from 6527 to 6594 \r{A}. 

Spectra for the 2005 to 2018 period were also obtained from the Lowell Observatory in Flagstaff, Arizona, USA. These were taken with the 1.1\,m John S. Hall telescope equipped with an echelle spectrograph with a resolving power of $R$ = 10,000. These observations from the Lowell Observatory were first published in \citet{suf20}. Three additional spectra for 2022 used in this study have been acquired at the 0.4\,m telescope of the Brooks Observatory at Central Michigan University (CMU). The telescope was equipped with LHIRES III spectrograph from Shelyak Instruments with 2400 lines/mm grating, used in conjunction with a Atik 460EX monochrome CCD camera and 50 $\mu$m slit. This resulted in H$\alpha$ spectra with a resolving power of $R$ = 11,000.

For the period spanning 2011 to 2023, additional observations were taken from the BeSS database\footnote{\href{http://basebe.obspm.fr/basebe/}{http://basebe.obspm.fr/basebe/}} \citep{nei11}, observed with a 0.4\,m telescope at the observatory of the Vereinigung der Sternfreunde K\"oln, near Cologne, Germany. For 2011-2017, these were taken with a Shelyak LHIRES III spectrograph with a resolving power of $R$ = 11,000. For 2017-2023, the spectra were taken with a Shelyak LHIRES III spectrograph with a resolving power of $R$ = 20,000. 

All spectral files for $\delta$ Sco containing the He\,\textsc{i} 6678 \r{A} line were also downloaded from BeSS. Spectra with low SNR or bad wavelength calibration were deleted to remove outliers, and selected observations were not restricted to a specific observing site. The available data provided coverage for the periastron in 2011, but not in 2000 or 2022.

\subsection{Polarimetry}

\vspace*{-2.5em}

\begin{deluxetable}{lccc}
\tabletypesize{\scriptsize}
\tablewidth{0pt} 
\tablecaption{$\delta$ Sco's observed field stars. \label{field_stars}}
\tablehead{
\colhead{Object} & \colhead{Spectral Class\tiny{\tablenotemark{a}}}& \colhead{Angular Separation ($^{\circ}$)} & \colhead{Distance\tiny{\tablenotemark{b}} (pc)}
} 
\startdata 
$\delta$ Sco & B0.3IV & 0 & $150\substack{+0.4 \\ -0.5}$ \\
HD 142902 & A9V & 0.54 & 149.5 $\pm$ 0.4\\
HD 143567 &  B9V &  0.74 & 134.7 $\pm$ 0.5 \\ 
HD 143600 &  B9.5V & 0.44 & 142.6 $\pm$ 0.6 \\ 
HD 144548 & F7V & 1.71 & 150.9 $\pm$ 1.3 \\ 
\enddata
\tablenotetext{a}{Spectral Class taken from SIMBAD database.}
\tablenotetext{b}{Distance for $\delta$ Sco from Hipparcos \citep{van07}; distances for the field stars based on Gaia DR3 \citep{bab23}}
\end{deluxetable}

\begin{deluxetable}{lcccc}
\tabletypesize{\scriptsize}
\tablewidth{0pt} 
\tablecaption{Parameters of the Serkowski function for the field stars of $\delta$ Sco and the mean polarization angles. \label{serk}}
\tablehead{
\colhead{Object} & \colhead{$P_{max}$ (\%)}& \colhead{$\lambda_{\rm max}$} & \colhead{K} & \colhead{$\left \langle \theta  \right \rangle$ (°)}
} 
\startdata 
HD 142902 & 0.411 $\substack{+0.084 \\-0.004}$ & 0.568 $\substack{+0.018\\-0.475}$ & 0.95 $\pm$ 0.07 & 148.0 $\pm$ 0.3  \\
HD 143567 & 0.553 $\substack{+0.008\\-0.007}$ & 0.595 $\substack{+0.016\\-0.535}$ & 1.00 $\pm$ 0.08 & 142.3 $\pm$ 0.3 \\ 
HD 143600 & 0.628 $\substack{+0.008\\-0.099}$ & 0.614 $\substack{+0.013\\-0.569}$ & 1.03 $\pm$ 0.08 & 144.6 $\pm$ 0.4 \\ 
HD 144548 &  0.602 $\substack{+0.008\\-0.008}$ & 0.703 $\substack{+0.010\\-0.011}$ & 1.18 $\pm$ 0.08 & 144.4 $\pm$ 0.6 \\ 
\enddata
\end{deluxetable}

The polarimetric observations of $\delta$ Sco were carried out beginning in 2010 at Pico dos Dias Observatory (OPD), located in Minas Gerais, Brazil, and operated by the National Laboratory of Astrophysics (LNA). 
These observations were conducted using the 1.60m Perkin-Elmer telescope and the 0.60m Boller \& Chivens telescope, jointly with the IAGPOL polarimeter (described by \citealp{mag96}) to obtain linear polarization in the optical range with {\it{BVRI}} filters. The polarimeter consists of a half-wave retarder and a calcite prism analyzer, with the light beams ultimately being recorded by a CCD camera. In a typical observation, a sequence of images at eight half-wave plate positions (WPP) separated by 22.5$^\circ$ is acquired. For each observing night, at least one polarized standard star was observed to calibrate the polarization angle in the equatorial system. Data reduction was performed using software packages described by \citeauthor{mag96} (\citeyear{mag84}, \citeyear{mag96}) and \cite{car07}. Data for $\delta$ Sco prior to 2010 were obtained from the MAST archive\footnote{\href{https://archive.stsci.edu/hpol/index.html}{https://archive.stsci.edu/hpol/index.html}}. These data were collected using the University of Winsconsin's Half-wave Spectropolarimeter (HPOL) at the Pine Bluff Observatory near Madison, Winsconsin \citep{nord96}.

In parallel with the polarimetric observations, additional observations were conducted at OPD to determine the interstellar polarization (ISP) along the line of sight of $\delta$ Sco, using the same observational procedures and data reduction methods previously mentioned. Knowing the value of the ISP is crucial for the study of Be stars because it allows us to ascertain the intrinsic polarization of the star, which arises due to the presence of the disk. The method of field stars is the simplest way to determine the value of ISP. It involves selecting and observing stars that are physically close to the target both in radial and angular distance. If these field stars do not exhibit intrinsic polarization, the observed polarization will serve as a proxy of the ISP. Four field stars of $\delta$ Sco were observed: HD 144548, HD 143567, HD 142902, and HD 143600. Additional data for these stars and $\delta$ Sco are detailed in Table \ref{field_stars}.

The spectral dependence of linear polarization in the interstellar medium with wavelength is described by the Serkowski Law \citep{ser75}
\begin{equation}
    \label{serkowski}
    \frac{P(\lambda)}{P_{\rm max}} = \exp \left[-K \ln^2 \left( \frac{\lambda_{\rm max}}{\lambda}\right)\right],
\end{equation}
where $P_{\rm max}$ is the maximum polarization that occurs at a wavelength $\lambda_{\rm max}$, $K$ is a dimensionless constant that describes the inverse width of the polarization curve peaking around $\lambda_{\rm max}$ \citep{cot19}. If $K$ is treated as a third free parameter, it can be shown that $\lambda_{\rm max}$ and $K$ are linearly related following the relationship (\citealt{mar92})
\begin{equation}
    \label{wilking}
    K=c_1 \lambda_{\rm max} + c_2.
\end{equation}
\noindent
\citet{whi92} define the values of $c_1 = 1.66 \pm 0.09$ and $c_2 = 0.01 \pm 0.05$, with $\lambda_{\rm max}$ in micrometers ($\mu \rm m$). Equation \ref{serkowski}, with $K$ given by Eq.~\ref{wilking}, is also known as the Serkowski-Wilking Law \citep{wil80}.

For each dataset obtained for $\delta$ Sco's field stars, the Serkowski function (Eq.~\ref{serkowski}) is fitted using the Monte Carlo method via Markov Chains. The fits return the values of $P_{\rm max}$ and $\lambda_{\rm max}$ for each curve (Table \ref{serk}).  
In addition to the parameters above, an additional parameter is necessary to fully described the ISP, namely the polarization position angle, $\theta_{IS}$.
 
\begin{figure*}[!t]
\begin{center}
\includegraphics[width=2 \columnwidth,angle=0]{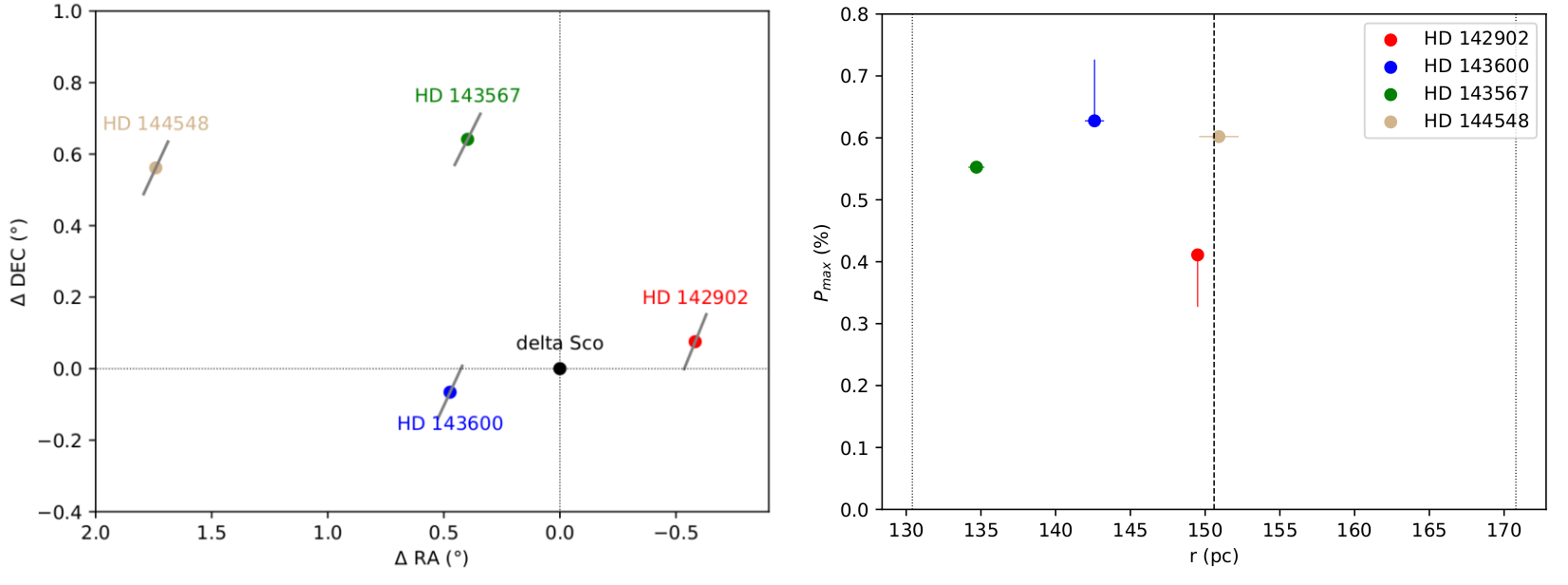}
\caption{Left: Polarization map for $\delta$ Sco's field. The direction of each bar represents the polarization position angle, while the length of the bars represents the magnitude of the maximum polarization. Right: Maximum polarization and distance of each field star. The dashed vertical line represents the distance of $\delta$ Sco, and the dotted lines represent the distance uncertainties.}
\label{PI_images}
\end{center}
\end{figure*}

Fig. ~\ref{PI_images} displays the positions of the stars in the sky (declination, $\Delta_{DEC}$, and right ascension, $\Delta_{RA}$) along with their respective polarization pseudovectors, representing the maximum polarization degree and polarization angle in each observed filter. In this map, there is consistency in the direction of the polarization pseudovectors, which may indicate that all the stars sample the interstellar medium behind which $\delta$ Sco is located. Fig. \ref{PI_images} (right) presents the maximum polarization of each field star along with its respective distance.

To find the maximum polarization of the interstellar medium ($P_{\rm max}^{\rm IS}$) and the interstellar polarization angle ($\theta_{IS}$), an average was calculated from the respective values of each field star, resulting in  $P_{\rm max}^{\rm IS} = 0.549\% \pm 0.003\%$ and $\theta_{IS}=144.8^{\circ} \pm 0.2^{\circ}$, respectively. In order to find the ISP in each filter, we used the Serkowski-Wilking Law, reaching the results $P_{\rm IS}^{\rm B}=0.490 \% \pm 0.006 \%$, $P_{\rm IS}^{\rm V}=0.539 \% \pm 0.003 \%$, $P_{\rm IS}^{\rm R}=0.549 \% \pm 0.003 \%$ and $P_{\rm IS}^{\rm I}=0.505 \% \pm 0.005 \%$, for the {\it{BVRI}} filters, respectively. 
\cite{hal58} found the polarization degree of $\delta$ Sco in $V$-band filter to be $P_{\rm V} = 0.60\%$, during the diskless phase of the star. Therefore, this value refers to the level of ISP along the line of sight. This value is very close to what we obtain from observations of $\delta$ Sco's field stars in the $V$-band ($P_{\rm IS}^{\rm V}=0.539 \% \pm 0.003 \%$).

\section{Results} \label{sec:results}

\begin{figure*}[htb!]
\epsscale{1.2}
\plotone{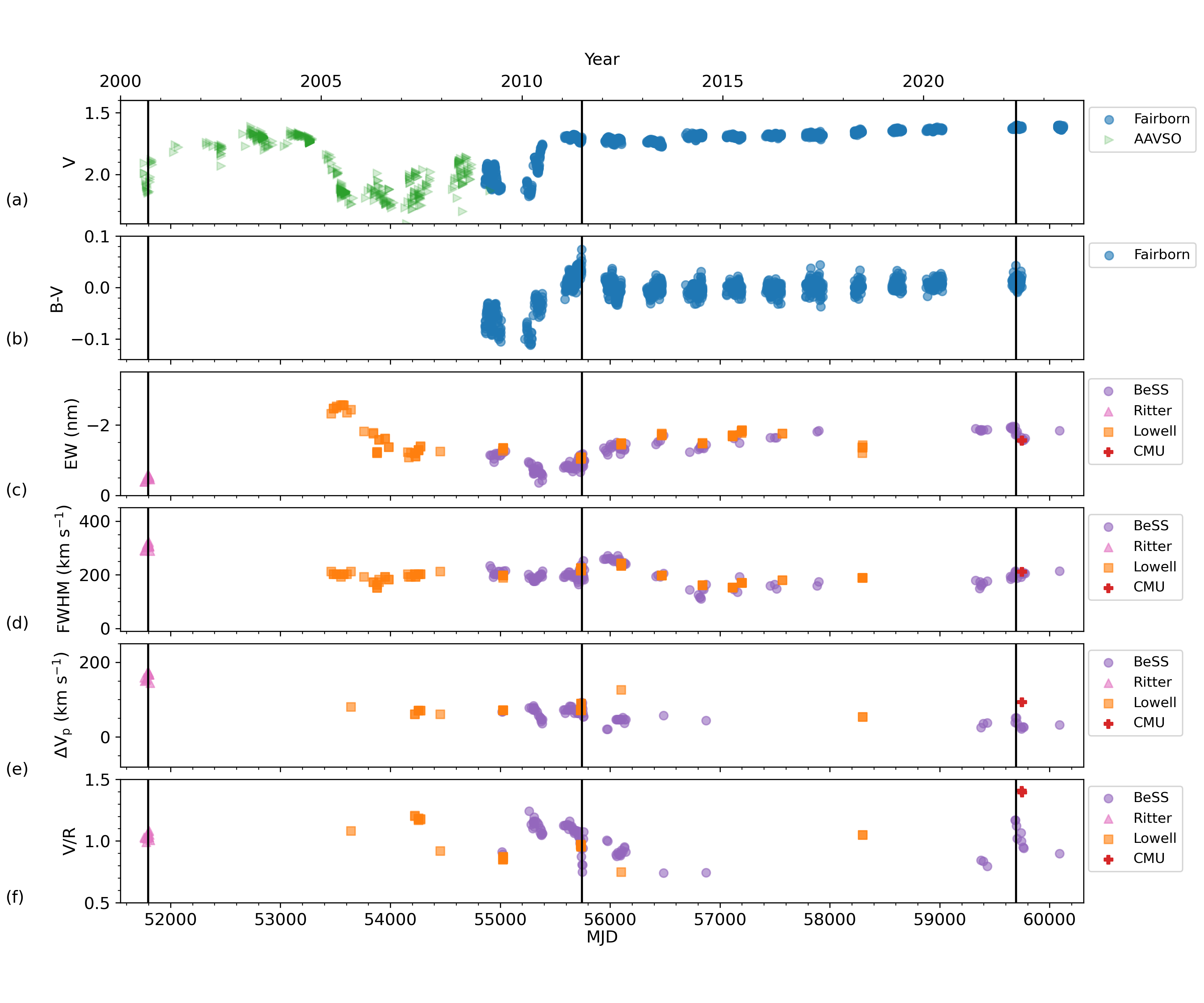}
\caption{$V$-band photometry and H$\alpha$ spectroscopy for $\delta$ Sco from 2000 to 2023. The time scale on the bottom x-axis is the Modified Julian Date (MJD), while the top x-axis provides the year. Panel a) shows $V$-band photometry, while panel b) shows the $B-V$ colour index and panels c), d), e), and f) show H$\alpha$ equivalent width (EW), full width at half maximum (FWHM), peak separation ($\Delta$v$_{p}$), and emission peak ratio (V/R), respectively. The legends indicate the data sources for each panel. The periastron times are indicated by vertical lines.}
\label{fig:longterm_comp}
\end{figure*}

\subsection{Photometry}

\begin{figure}[htb!]
\epsscale{1}
\plotone{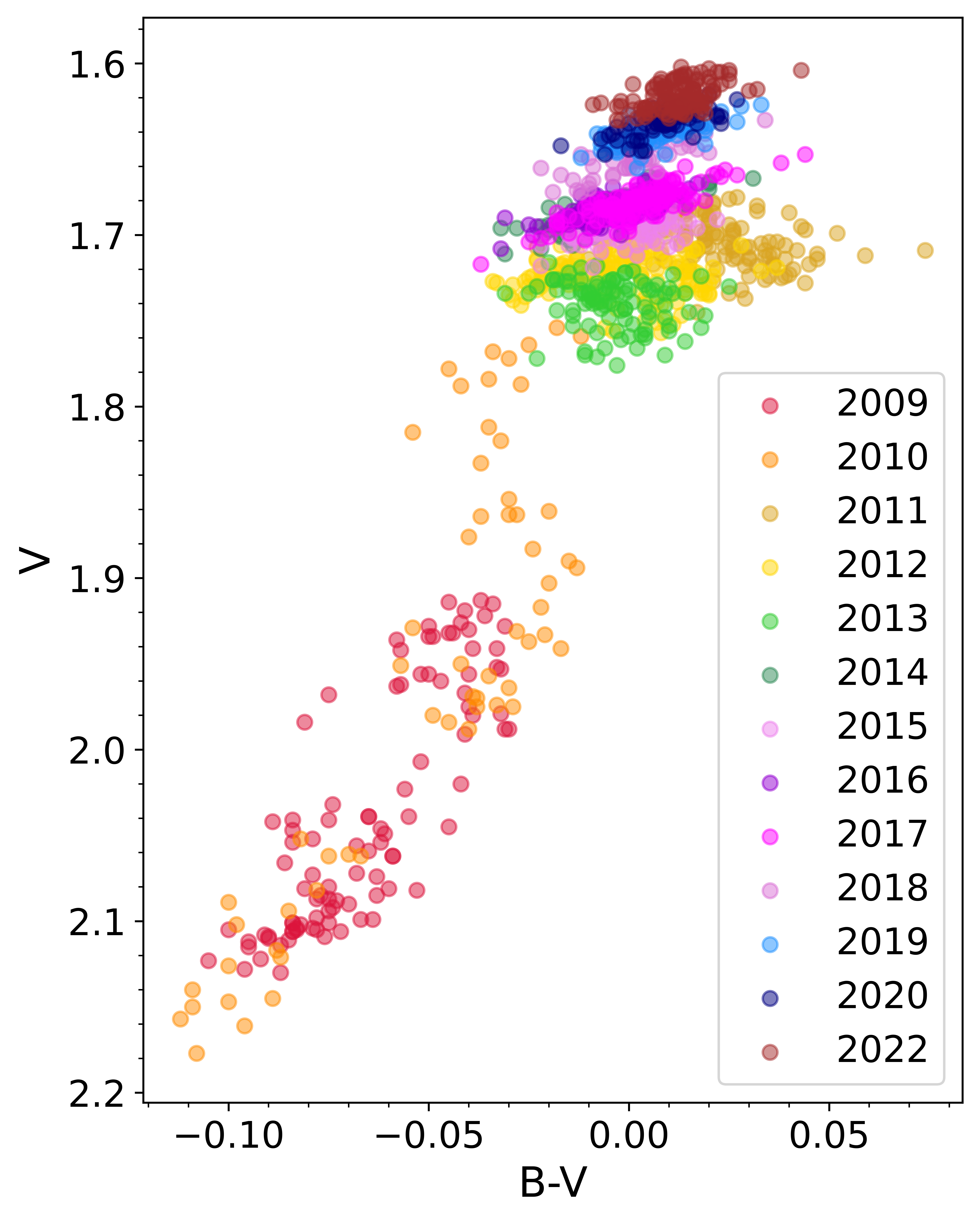}
\caption{$V$ magnitude vs. the $B-V$ colour index for $\delta$ Sco from 2009 to 2022. The legend indicates the data for each year.}
\label{fig:bminusv}
\end{figure}

$V$-band photometry for the years spanning 2000 to 2023 is shown in panel (a) of Fig. \ref{fig:longterm_comp}. The mean magnitude has brightened between each consecutive periastron, with the changes between 2000-2011 being especially large. The $B-V$ colour index, plotted in panel (b) of Fig. \ref{fig:longterm_comp}, shows variations that are concurrent with the $V$-band fluctuations, and they follow a similar trend. Figure \ref{fig:bminusv} shows the $V$ magnitude versus colour index ($B-V$) for 2009 to 2022. The system reddened during the time leading up to the 2011 periastron, but remained approximately the same colour between 2011 and 2022. 

\begin{figure*}[htb!]
\plotone{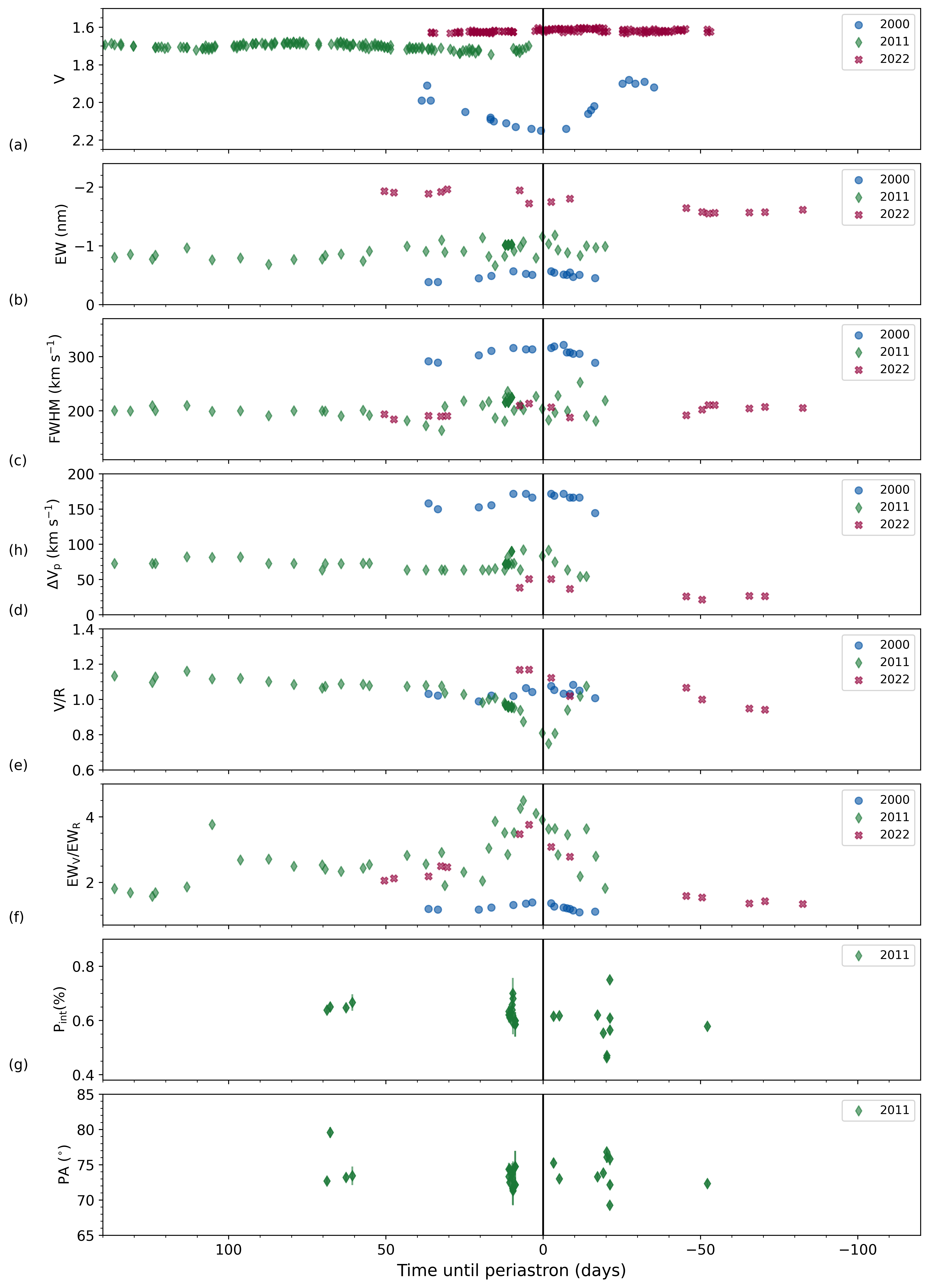}
\caption{Comparison of the $V$-band, H$\alpha$ and polarization data for $\delta$ Sco during the time surrounding the 2000, 2011 and 2022 periastra. Panel a) shows the $V$-band magnitude. Panels b), c) d), e) and f) show the equivalent width (EW), full width at half maximum (FWHM), peak separation ($\Delta$v$_p$), emission peak ratio (V/R), and EW$\mathrm{_V}$/EW$\mathrm{_R}$ for the H$\alpha$ line, respectively. Panel (g) shows the $V$-band intrinsic polarization, while panel (h) shows the PA. The x-axis measures time until the periastron for each year. The vertical lines represent the periastra.}
\label{fig:periastra_obs}
\end{figure*}

Panel a) of Fig. \ref{fig:periastra_obs} shows the $V$-band data for the three most recent periastra. The system dimmed by a small amount around the 2000 periastron, but this may have been related to the large-amplitude variability observed in the following years. \citet{jon11} found cyclical brightness variations in the data spanning 2009 to 2012, although these cycles have not been detected in the photometry from 2013 onwards. The data do not indicate significant perturbations of the disk in 2011 or 2022.

\subsection{Spectroscopy}

As mentioned in Section \ref{sec:intro}, previous studies on $\delta$ Sco have been unable to detect the spectral lines of the secondary star. If the radial velocities between the primary and secondary stars at periastron are sufficiently large, it could be possible to differentiate the lines originating from the primary and the secondary. However, the maximum radial velocity of the primary star at periastron is $\approx$ 50 km s$^{-1}$, corresponding to a shift of roughly 0.1 nm \citep{mir13, tan09}. These shifts are small compared to the width of the H$\alpha$ line, and therefore the secondary star remains undetected through spectroscopy.

Panels (c), (d), (e) and (f) of Fig. \ref{fig:longterm_comp} show the EW, FWHM, peak separation $\Delta$v$_p$ and V/R of the H$\alpha$ emission line from 2000 to 2023. Overall, the data during the periastra do not show abnormalities in comparison to the rest of the observations. The V/R and FWHM variations in 2011 are the main exceptions to this trend.

\begin{figure}[htb!]
\epsscale{1.25}
\plotone{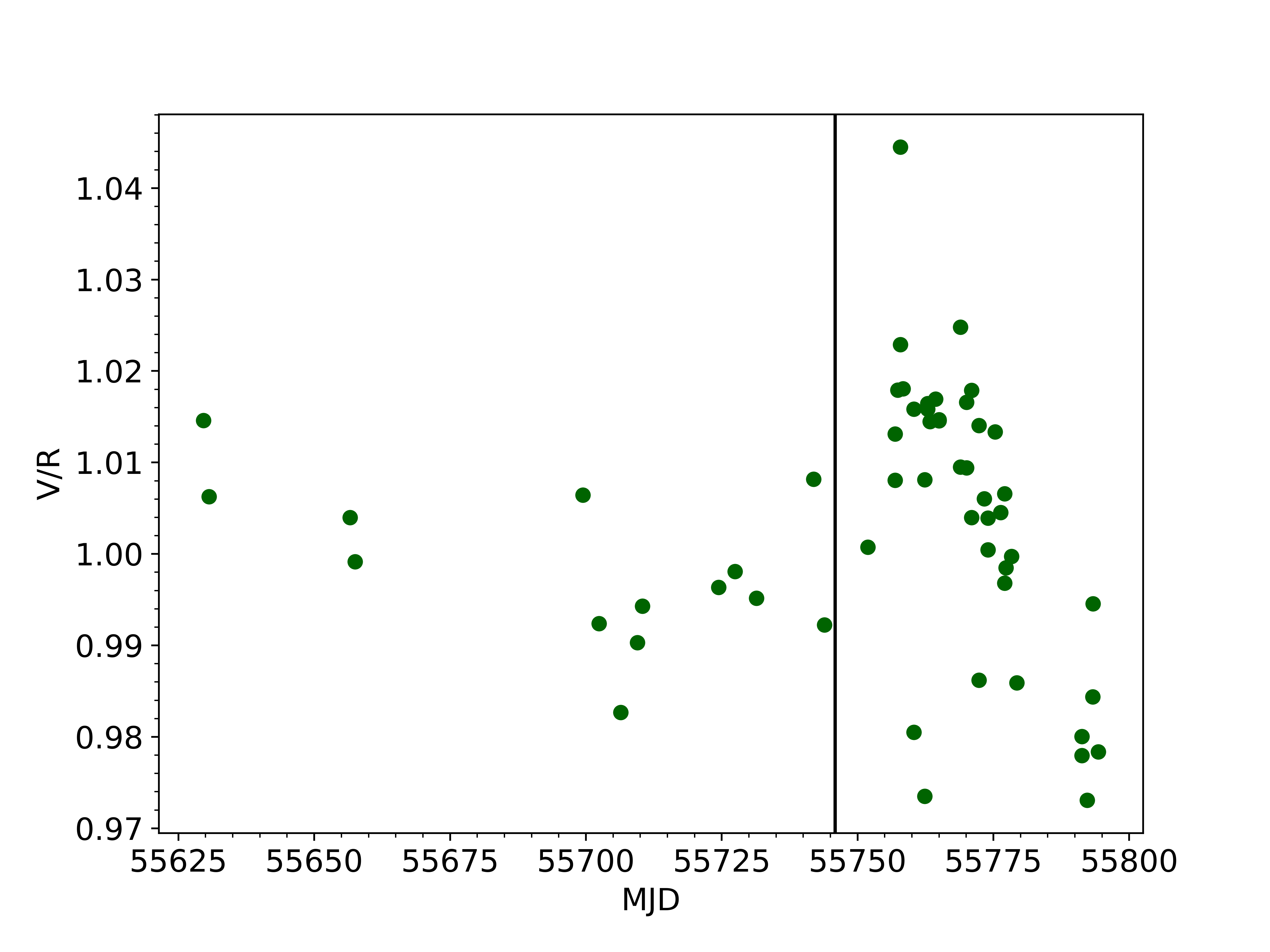}
\caption{Emission peak ratio (V/R) for the He\,\textsc{i} 6678 line during 2011. The vertical line represents the periastron time.}
\label{fig:he_lines}
\end{figure}

Fig. \ref{fig:he_lines} shows the He\,\textsc{i} 6678 \r{A} line for the time period surrounding the 2011 periastron. Similar to the H$\alpha$ line, this spectral line shows enhanced blue emission near periastron, but delayed by approximately one month. Since the He\,\textsc{i} 6678 line forms in the inner disk while the H$\alpha$ line forms over a larger emitting volume, this lag may indicate that the outer disk was perturbed first, then the helium emitting volume was perturbed later.

Panels (b), (c), (d) and (e) of Fig. \ref{fig:periastra_obs} plot the H$\alpha$ EW, FWHM, $\Delta$v$_p$, and V/R for only the periastra. Panel (f) of  Fig. \ref{fig:periastra_obs} plots the ratio of the EW of the violet side of the H$\alpha$ line to the EW of the red side (EW$\rm _V$/EW$\rm _R$). The radial velocities for the H$\alpha$ line were found using the mirrored profile method, fitting the wings of the lines to compare the direct and reversed line profiles. The radial velocity measurement for each observation was then used as the line center, and the EW of the line on either side of this value was calculated. This ratio can be used to trace line asymmetries when the profile is singly-peaked or shows complex peak structure, in which cases the V/R ratio cannot be used.

Fig. \ref{fig:periastra_obs} shows a general lack of repetitive patterns throughout the periastra. With the exception of the V/R ratio, the measured values leading up to each periastron are different. The unifying feature between periastron years is the EW$\rm _V$/EW$\rm _R$, which consistently peaks near periastron.

\subsection{Polarimetry}

\begin{figure*}[htb!]
\plotone{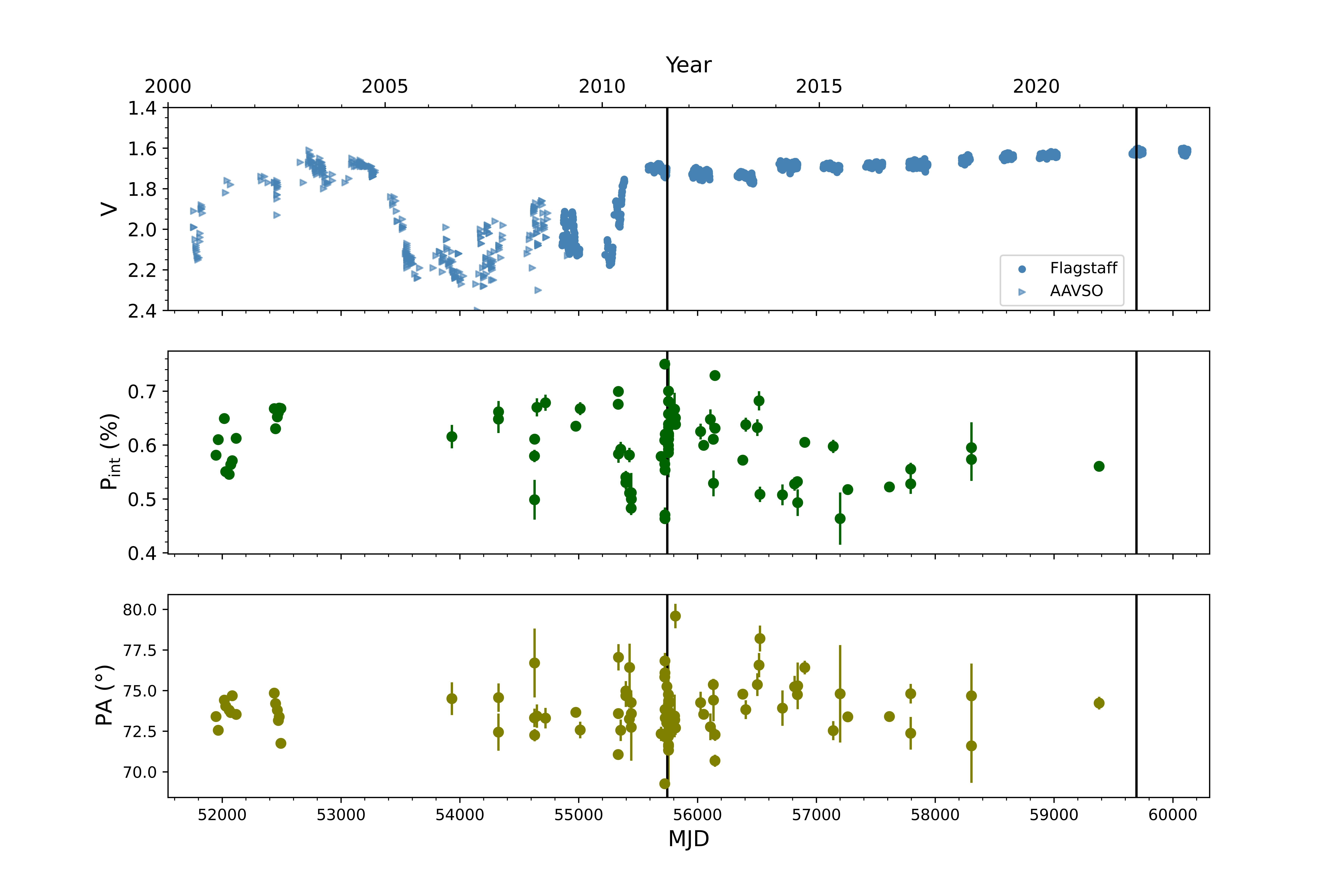}
\caption{Top: $V$-band photometry for $\delta$ Sco during the period 2001-2023. Center: Intrinsic polarization degree. Bottom: Polarization angle. The time scale on the bottom x-axis is Modified Julian Date, while the scale on the top x-axis is the decimal year. 
\label{fig:polarization_longterm}}
\end{figure*}

By vector subtraction, we can extract the intrinsic data of $\delta$ Sco by separating the interstellar component from the observed measurements. Fig. \ref{fig:polarization_longterm} exhibits the variation of intrinsic polarization degree and angle over time for the $V$-band. We computed an average of the intrinsic polarization angles in all four filters, resulting in $\theta = 73.42^{\circ} \pm 0.14^{\circ}$.

An alternative method for acquiring the intrinsic polarization angle of a star is through the observed $Q$ and $U$ Stokes parameters \citep{dra14}. The positioning of $\delta$ Sco's data points on the $QU$ diagram for the $V$-band is illustrated in Fig.~\ref{QU_diagram}. The $QU$ diagrams for the remaining filters are found in the Appendix. The data in all filters adhere to a consistent linear pattern, signifying a steady intrinsic angle of $\theta=74.4^{\circ} \pm 0.1^{\circ}$. This angle is in agreement with the calculated intrinsic measurement obtained using the field stars method, and offers strong supporting evidence for the ISP determination.

\begin{figure*}
\gridline{\fig{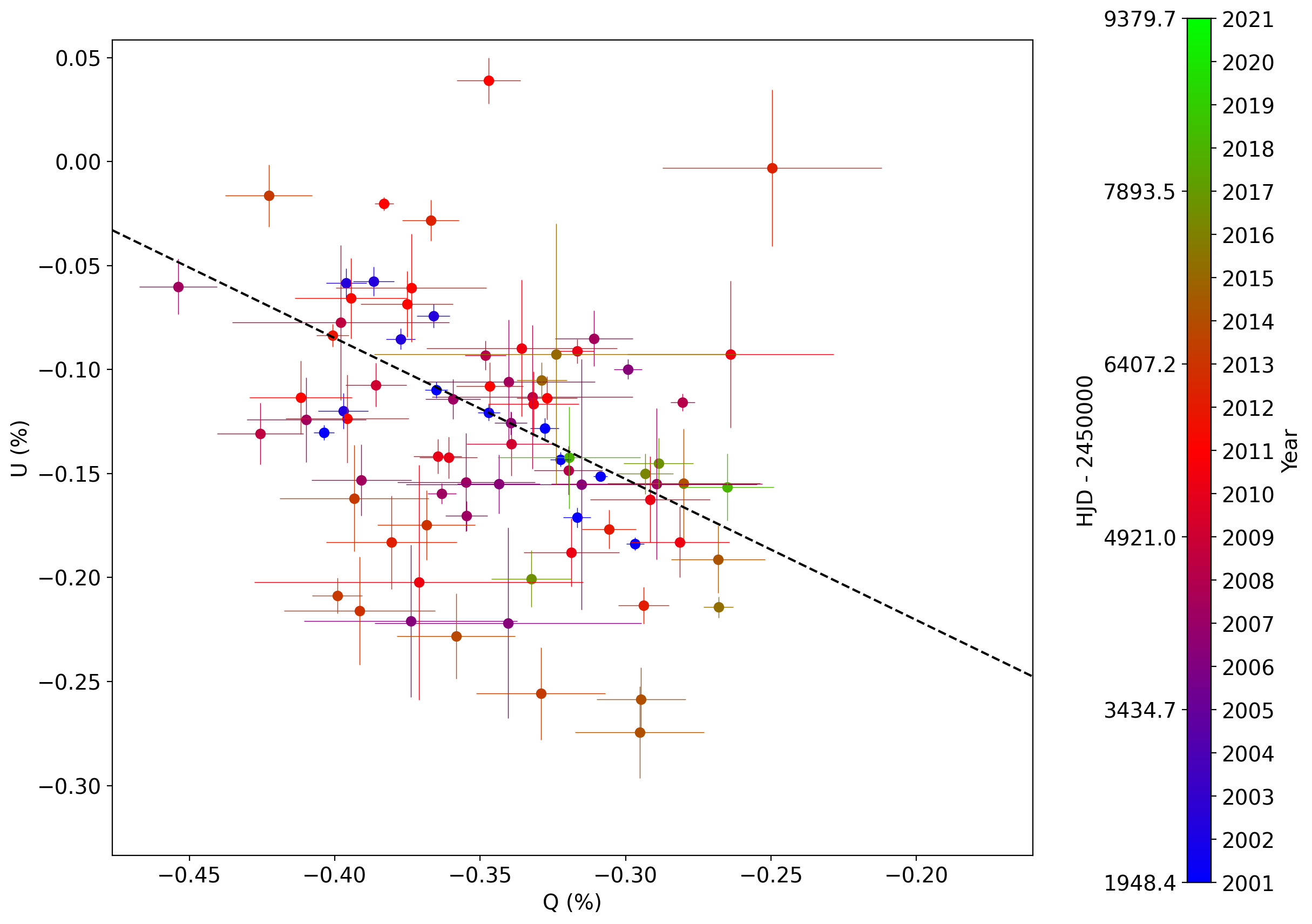}{0.5\textwidth}{(a)}
          \fig{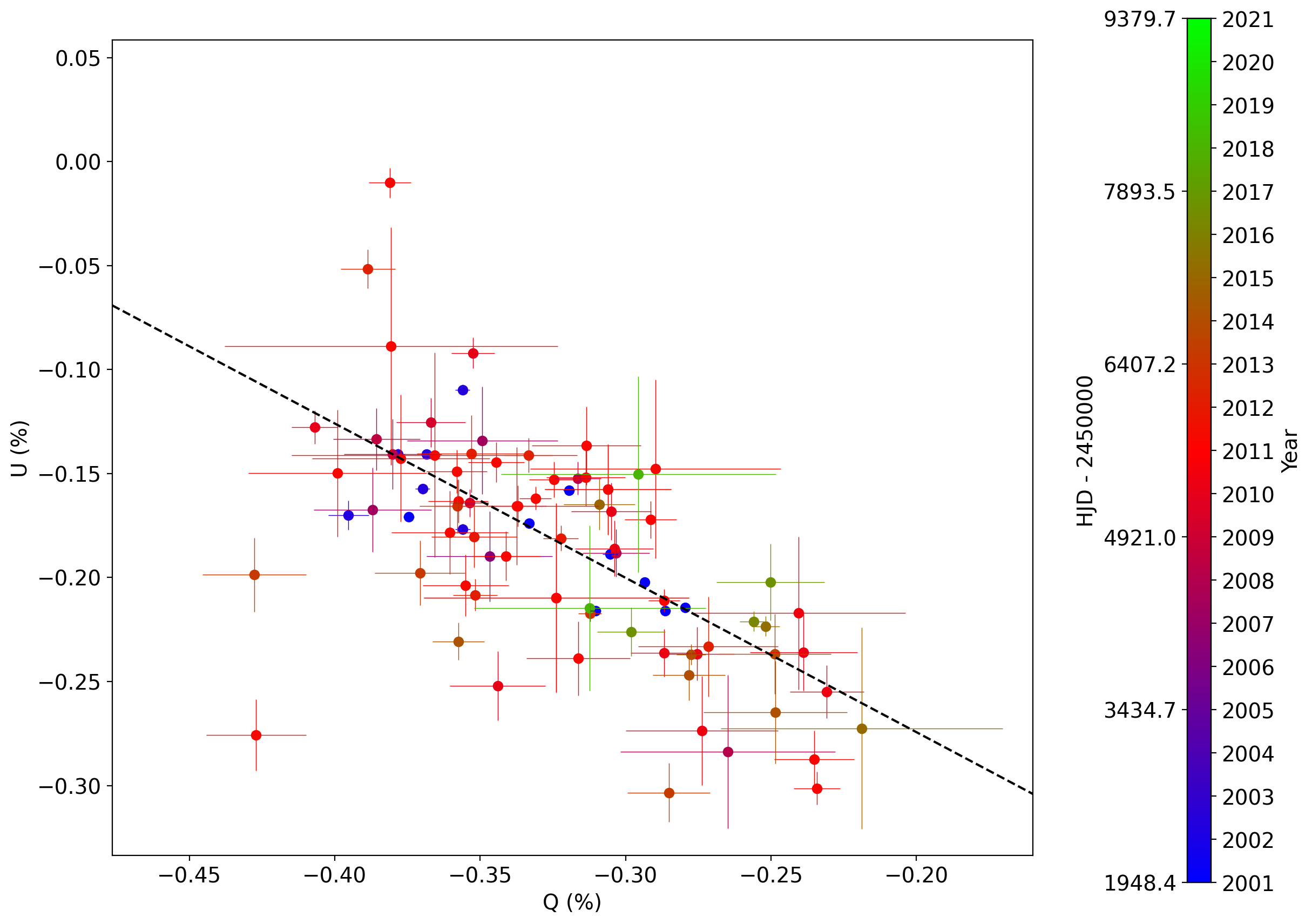}{0.5\textwidth}{(b)}}
\gridline{\fig{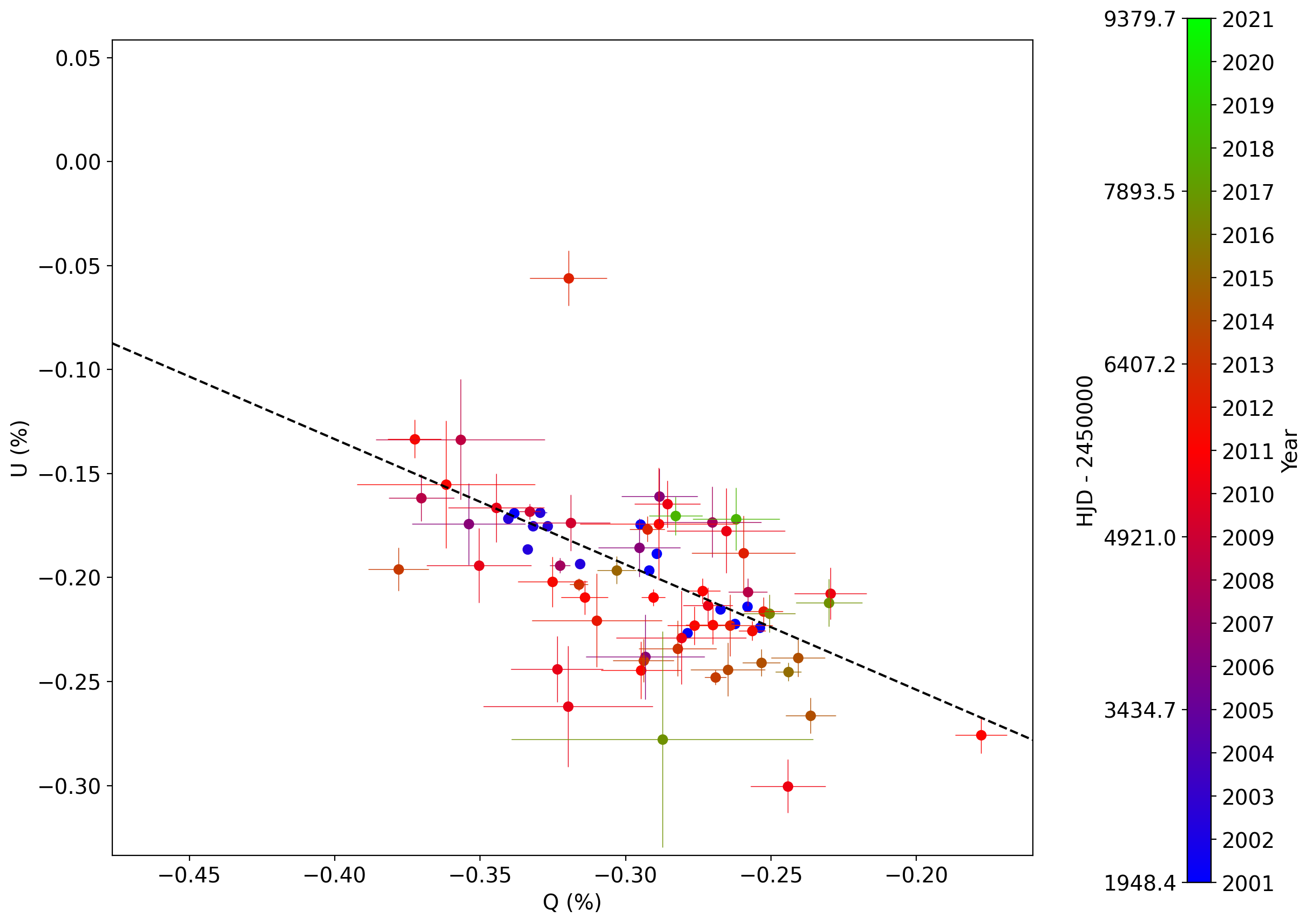}{0.5\textwidth}{(c)}
          \fig{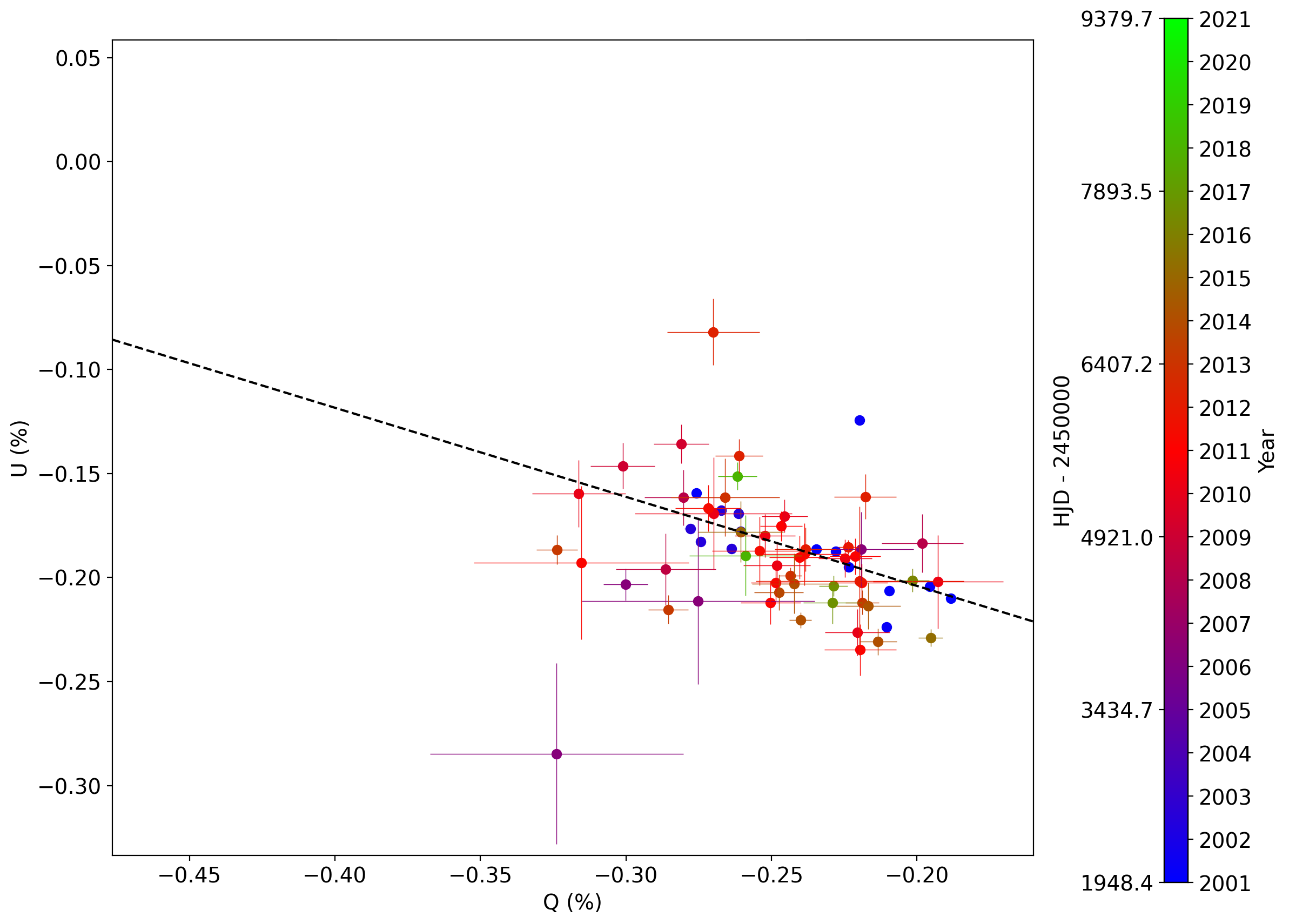}{0.5\textwidth}{(d)}}
\caption{(a) QU Diagram of $\delta$ Sco in the $B$-band. The dashed lines indicate the most likely solution of the damped least-squares method (Levenberg–Marquardt algorithm), with a reduced chi-squared ($\chi^2$) of 31.8678. (b) Same as in (a), but for the $V$-band, with reduced $\chi^2=24.9883$. (c) Same as in (a), but for the $R$-band, with reduced $\chi^2=22.0285$. (d) Same as in (a), but for the $I$-band, with reduced $\chi^2=56.4171$}. 
\label{QU_diagram}
\end{figure*}

Fig. \ref{fig:polarization_longterm} compares the long-term trends in polarization vs. $V$-band photometry. The polarization degree, shown in panel (b), shows a gradual increase leading up to 2011, followed by a decline over the next four years. By 2015, the polarization degree stabilized at its lowest recorded values.  Curiously, this minimum in the average value of the polarization corresponds to when the V-band photometric data is close to its maximum value. The position angle (PA) is shown in panel (c). This shows the greatest range of variation in 2011, but remains roughly stable in every other year. The scatter observed in the data is not a result of observational errors; instead, it represents intrinsic short-term variations.

\section{Discussion} \label{sec:discussion}

The highly eccentric binary orbit of $\delta$ Sco has important implications for the tidal interactions between the secondary star and the primary’s disk. The periastron passage is a brief event; if we assume the orbital parameters found by \citet{tyc11}, the secondary is within 50 $R_{\star}$ of the primary for no more than 40 days. Here, we choose the astrometric solution for parallax rather than the dynamical parallax estimates by \citet{mir13}, \citet{tyc11} and \citet{tan09}, due to the large flux ratios of the system. Although the minimum separation of the two stars is small (approximately 29 $R_{\star}$, assuming a stellar radius of 6.8 R$_{\odot}$), the secondary star is acting on the disk at this distance for only a matter of hours. Since the secondary star is near the primary for a small portion of its orbit, the interaction time is much smaller than for systems with smaller eccentricities \citep{pan16}. For the remainder of the system’s nearly 11-year period, the disk is not influenced by the secondary star and can evolve independently. At each encounter, the secondary star interacts with a disk with different characteristics. 

The observed trends in photometry, spectroscopy and polarimetry give useful insight into how the disk characteristics differ leading up to each closest approach, resulting in distinct patterns in the observations for each passage. We will examine the photometric evidence first, then turn to the H$\alpha$ spectra and polarimetry. We then conclude this section with a summary comparing the physical characteristics of the disk at each periastron passage.

\subsection{Disk evolution}

After the periastron in 2000, the $V$-band photometry shown in panel (a) of Fig. \ref{fig:longterm_comp} shows a significant drop in brightness beginning in 2005 which reached a minimum between 2006 and 2007. The H$\alpha$ EW, shown in panel (c) of Fig. \ref{fig:longterm_comp}, exhibits a similar pattern during these years, although its response lagged behind the $V$-band because the H$\alpha$ emission line is formed in a larger volume of the disk \citep{car11}. \citet{suf20} discussed these variations in detail, attributing the drop in brightness and H$\alpha$ emission strength to partial dissipation of the young disk. Such behaviour has also been observed in 66 Oph \citep{mar21}, as well as $\omega$ CMa \citep{gho18}.

The trend in $V$ vs. $B-V$ shown in Fig. \ref{fig:bminusv}, and the long-term trends shown in panels (a) and (b) in Fig. \ref{fig:longterm_comp}, indicate that $\delta$ Sco has been both brightening and reddening since 2009. For (near) pole-on viewing, as is the case for this star, this reddening is expected during periods of disk growth \citep{hau12}. In simple terms, reddening occurs because the disk's contribution to the total flux builds as the disk grows \citep{vie17}, and the disk has a lower temperature than the star \citep{car06a, vie15}. Most of the growth appears to have occurred between 2009 and 2012, as those years correspond to the largest changes in both $V$ and $B-V$. While the system brightened by roughly one-tenth of a magnitude in the $V$-band between the two most recent periastron passages, the system did not redden significantly during this time. Fig. \ref{fig:bminusv} is a standard example of the variability resulting from the growth and dissipation of a disk for a pole-on Be star, as predicted by \citet{hau12} and observed in numerous Be stars (for example, see \citealt{dew06}).

The decrease in FWHM values between 2000 and 2011, as shown in panel (c) of Fig. \ref{fig:periastra_obs}, also points to a period of disk growth during the same period. This is consistent with the findings of \citet{suf20}, who concluded on the basis of H$\alpha$ EW values and $V$-band photometry that the disk underwent disk building and dissipating periods between 2000 to 2011, yielding an overall increase in the spatial extent of the disk during this period. The leveling off of FWHM values after 2011 suggests that disk growth was less significant between 2011 and 2022. However, the disk must have grown somewhat, evidenced by the decrease in peak separation between 2011 and 2022 and the slow secular increase in $V$-band flux in the past decade.

The radius of the disk, $R_d$, can be related to the peak separation $\Delta$v$_p$ and rotational velocity of the star $v \sin{i}$ through 
\begin{equation}
    \frac{R_d}{R_{\star}} = \left( \frac{2 v \sin{i}}{\Delta v_p} \right)^2 \,,
\end{equation}
assuming Keplerian rotation \citep{hua72, hum95, mei07}. Adopting a v $\sin{i}$ of 175 $\kmps$ \citep{ber70}, then, we can use H$\alpha$ peak separation values to roughly estimate the radius of the H$\alpha$ emitting region (ER) during each periastron year. 

In 2000, peak separation values began to change roughly 20 days before periastron. They rose from 150 $\kmps$, corresponding to an H$\alpha$ ER of roughly 5 $R_{\star}$, to approximately 170 $\kmps$, corresponding to an H$\alpha$ ER of 4.2 $R_{\star}$ at closest approach. These values for peak separation at periastron are slightly larger than those reported by \citet{mir01}, but yield values for the H$\alpha$ ER radius that are consistent with the disk radius of 5.3 $R_{\star}$ reported by \citet{mir03} for 2003, and somewhat lower than the estimate of 7 $R_{\star}$ by \citet{car06b}. All estimates agree that the H$\alpha$ ER radius in 2000 fell well within the first Lagrange point at $\approx$ 20 $R_{\star}$, \citep{mir13, egg83, kop59}, so mass transfer between the primary and the secondary star during periastron during the first periastron, in 2000, is unlikely. The secondary star would have passed no closer than 19 to 22 $R_{\star}$ from the H$\alpha$ ER of the disk.

In 2011, the peak separation had lowered to approximately 70 $\kmps$, corresponding to an H$\alpha$ ER radius of about 25 $R_{\star}$. This value is similar to the model-based estimate for the size of the H$\alpha$ ER by \citet{suf20} for that year, but somewhat larger than the interferometry measurements of 14.9 $R_{\star}$ in 2007 and 9 $\pm$ 3 $R_{\star}$ in 2010 by \citet{mil10} and \citet{mei11}, respectively. If we ignore any truncation effects on the disk, the secondary star could have passed within a few $R_{\star}$ of the H$\alpha$ ER. However, the peak separation reached a maximum of roughly 90 $\kmps$, which indicates the H$\alpha$ ER was truncated to approximately 15 $R_{\star}$ at closest approach.

\begin{figure*}[htb!]
\epsscale{1}
\plotone{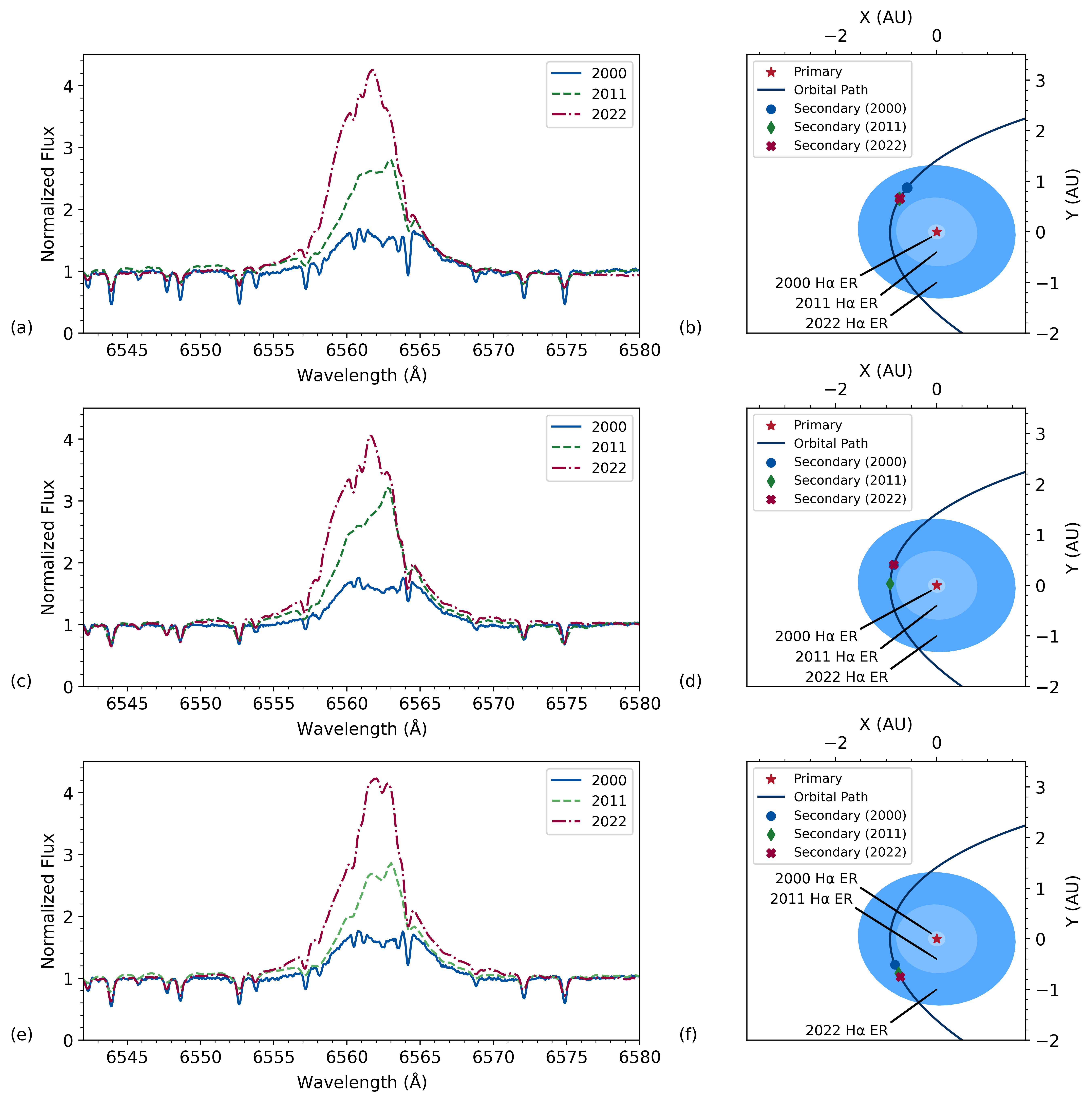}
\caption{a) The H$\alpha$ line profile as observed approximately one week before closest approach in 2000, 2011 and 2022. The 2000 spectrum was taken nine days prior, while the 2011 and 2022 spectra were captured seven days before closest approach. The narrow absorption features in the continuum are telluric lines. (b) A top-down view of the position of the secondary star relative to the primary star and the H$\alpha$ emitting region (ER) at the time each spectrum in (a) was observed. Sizes of the primary and secondary stars are not to scale. (c) Same as in (a), except the selected spectra were observed on the dates nearest closest approach. The spectrum in 2000 was observed three days prior to closest approach, the spectrum for 2011 was observed on the day of closest approach, and the spectrum for 2022 was taken four days before closest approach. (d)  Same as in (b), but for (c). (e) Same as in (a), except the selected spectra were observed approximately one week after closest approach. The spectrum in 2000 was observed ten days after, while the spectrum for 2011 was observed eight days after and the spectrum for 2022 was taken nine days after closest approach. (f) Same as in (b), but for (e).
\label{fig:spectrum_position}}
\end{figure*}

In 2022, the double-peaked structure of the H$\alpha$ line, as seen in panel (a) of Fig. \ref{fig:spectrum_position}, was not discernible until the week before minimum separation between the primary and the secondary. Peak separation at closest approach was approximately 50 $\kmps$, corresponding to a disk radius of 49 $R_{\star}$. This value is consistent with the estimate by \citet{suf20} for 2018. Interestingly, inspection of the changes in the line profile in 2022 reveals that the upper limit for the disk radius may be smaller than indicated by the peak separation. The line profile remains relatively unchanging and singly peaked up to 30 days before closest approach. After this point, the line shows a more complicated profile with significant asymmetries and increased peak separation indicating partial disk destruction. If these asymmetries are caused by interactions with the secondary star, this may suggest that the upper limit for the disk's H$\alpha$ ER in early 2022 is roughly 30 $R_{\star}$, since line profile changes were not apparent until the secondary star was within this distance.

\begin{figure*}[htb!]
\plotone{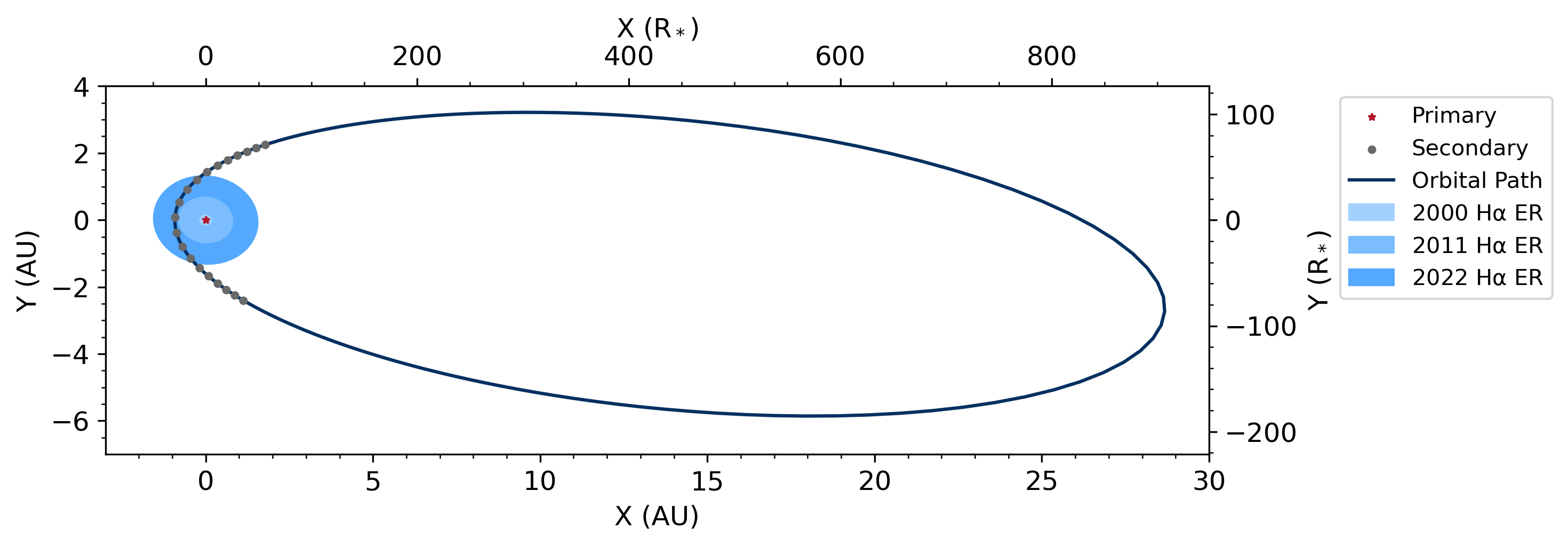}
\caption{The binary orbit of $\delta$ Sco, in the orbital x-y plane, showing the orbital solution (blue line) relative to the primary star (red star) along with the estimated sizes of the disk's H$\alpha$ emitting region (ER) for each periastron year (concentric blue circles). The position of the secondary star (gray circles) is plotted at 5 day intervals for 50 days before and after minimum separation. The motion of the secondary star is counter-clockwise.
\label{fig:orbital_diagram}}
\end{figure*}

Fig. \ref{fig:orbital_diagram} shows the binary orbit of $\delta$ Sco on the orbital x-y plane, where the origin is centered on the primary star. The orbital path of the secondary star was calculated using the orbital parameters published in \citet{tyc11}, which were produced from high angular resolution interferometric observations. The extent of the H$\alpha$ ER for each periastron year, based on the peak separation values discussed above, is also plotted. The secondary star, whose position is plotted at 5-day intervals for 50 days prior and following periastron, interacts with the disk for different lengths of time during each consecutive periastra. 

Panel (a) of Fig. \ref{fig:spectrum_position} compares the observed H$\alpha$ line profile roughly 1 week before closest approach in each periastron year. Since these spectra were taken at approximately the same orbital phase (within 2 days of each other), the radial velocities of the system for each observation are comparable. The calculated position of the secondary star on its orbital path at the moment each spectrum was observed is shown in panel (b). The intensity of the line has become progressively larger for each periastron, indicative of disk evolution during the intervening years. In 2000, the disk was the smallest and therefore the least perturbed of the three periastra, evidenced both by the larger peak separation and roughly equal peak heights. The profiles for 2011 and 2022, where the emission is stronger at smaller wavelengths, may  represent increasingly perturbed disks. In addition to showing higher relative flux at the central wavelengths, the relative flux of the violet side of the line increased significantly more than the red side for the two most recent periastron years. 

These asymmetries of the line profiles are still more apparent in the spectra observed nearest each periastron event, shown in panel (c) of Fig. \ref{fig:spectrum_position}. The 2011 spectrum shows a red peak enhancement that is not as pronounced in either of the other two years. The 2022 spectrum does not show much variation from panel (a), however this is not unexpected as the two spectra were observed only three days apart. Ratios of the EW of the violet side of the line to the EW of the red side (EW$\rm _V$/EW$\rm _R$) should demonstrate these asymmetries more clearly than the V/R ratios, due to the complex nature of the H$\alpha$ peak structure. Indeed, panel (f) of  Fig. \ref{fig:periastra_obs} shows the expected peaks in EW$\rm _V$/EW$\rm _R$ around closest approach. In the spectra observed roughly one week after periastron, presented in panel (e), a less complex double-peaked structure is more apparent in both 2011 and 2022.

We note that the H$\alpha$ line and the He\,\textsc{i} 6678 line both showed enhanced blue emission near the 2011 periastron, with the peak in the He I 6678 V/R observed approximately one month after the peak in the H$\alpha$ EWV/EWR. This excess blue emission could be caused by density and velocity perturbations propagating inward through the disk in response to the secondary star’s gravitational influence. As the secondary star approaches the primary during the time leading up to periastron, it interacts with the blue-shifted side of the disk first (see Figs. 2 and 3 in \citealt{che12}). Another contributing factor could be the secondary star illuminating the disk, heating the portion of it nearest the secondary which results in stronger line emission. This phenomenon has been observed before in the Be star 59 Cyg, where the secondary is a hot subdwarf star \citep{pet13}. In the future, a more dedicated modelling effort is needed to test these hypotheses.

Overall, the general lack of repetitive features in the data shown in Fig. \ref{fig:periastra_obs} can be explained by different disk characteristics at each passage of the secondary star. Due to the high eccentricity of the system, the Be disk can evolve independently of the secondary star for the majority of the orbit. This explanation is supported by radius estimates of the H$\alpha$ peak separations presented here, in addition to the long-term modelling of the system by \citet{suf20}. The asymmetric H$\alpha$ profiles during the close encounters in 2011 and 2022 indicate more significant disk perturbations during those events than during the 2000 passage when the disk was smaller.

\subsection{Polarization}

Panel (b) of Fig. \ref{fig:polarization_longterm} shows distinct drops in the $V$-band polarization degree in 2008 and 2010. These abrupt decreases may be explained by mass ejection events. When the mass injection rate increases suddenly, enhanced densities in the inner disk lead to increased continuum emission, which in turn causes the $V$-band polarization to drop \citep{hau14}. This downturn is temporary because, as the disk grows, the number of scatterers increases so the polarization must also rise \citep{hau14}. This would be consistent with Sco's behaviour in 2011, where the polarization rapidly increases to its highest levels in our observation set.  

Following the 2011 periastron, the drop in $V$-band polarization degree, rise in $V$-band photometry and decrease in H$\alpha$ EW shown in Fig. \ref{fig:longterm_comp} indicate sustained disk growth. The decrease in polarization degree could be explained through a combination of multiple scattering and post-scattering absorption. In a tenuous disk, polarization is initially linearly proportional to density. However, for large density values, the polarization begins to flatten out and eventually, it should begin to decrease. This is expected because photons scattered in dense disks may be subsequently reabsorbed rather than escaping the disk, causing an overall reduction in polarization degree.

\begin{figure}[htb!]
\epsscale{1.1}
\plotone{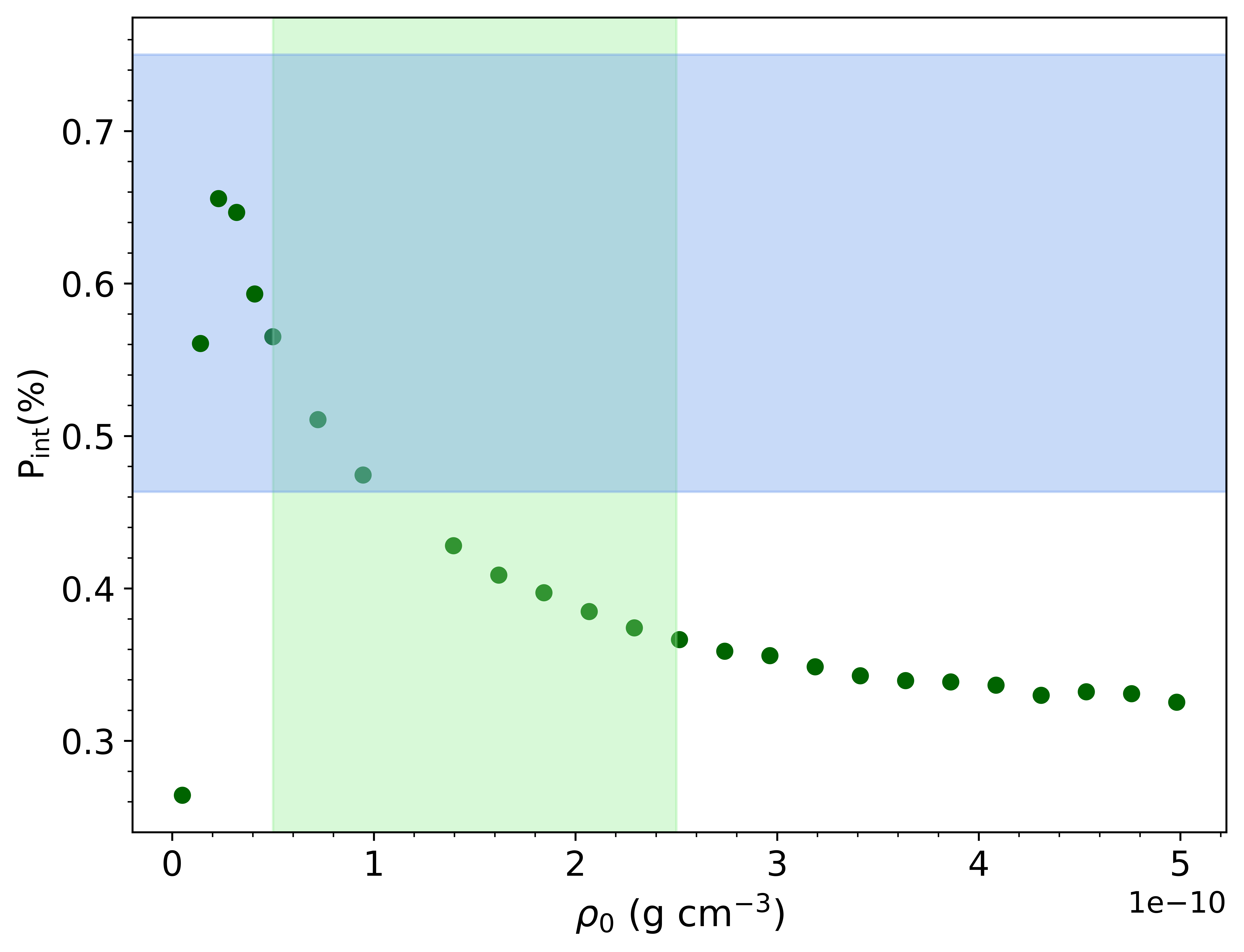}
\caption{Mean polarization degree ($P_{int}$) vs. the density at the innermost regions of the disk ($\rho_0$) for the $V$-band, produced by \textsc{HDUST} for an inclination angle of 35$^{\circ}$. The best-fit $\rho_0$ values for $\delta$ Sco from \citet{suf20} for 2010 to 2018 are shaded in light green. The light blue shaded region represents the observed intrinsic polarization in the $V$ band from 2010 to 2018.} 
\label{fig:HDUST_pol}
\end{figure}

To investigate this explanation, a set of 26 \textsc{HDUST} models were run with varying density values and with the stellar parameters listed in Table \ref{sph_params}. The number density $n_0$ of the disk was increased incrementally from 5$\times 10^{12}$ cm$^{-3}$ to 5$\times 10^{14}$ cm$^{-3}$. For these values of $n_0$, the density at the innermost regions of the disk $\rho_0$ has values of 5$\times 10^{-12}$ g cm$^{-3}$ to 5$\times 10^{-10}$ g cm$^{-3}$. \citet{suf20} found that the disk of $\delta$ Sco had $\rho_0$ $\approx$ 5$\times 10^{-11}$ g cm$^{-3}$ to 2.5$\times 10^{-10}$ g cm$^{-3}$ for 2010 to 2018. Fig \ref{fig:HDUST_pol} shows the mean polarization level vs. $\rho_0$ for these models. The $V$-band polarization degree was estimated by calculating the mean polarization degree across the $V$-band (500 to 650 nm). 

Fig. \ref{fig:HDUST_pol} shows that as $\rho_0$ increases, the intrinsic polarization degree also increases as expected. However, the intrinsic polarization degree reaches a maximum and then drops off for larger values of $\rho_0$ when the disk is very dense and optically thick. After $\rho_{0} \approx$ 3$\times 10^{-10}$ g cm$^{-3}$, the polarization degree stays approximately constant. Fig. \ref{fig:HDUST_pol} was constructed to help us understand the observed trends in polarization degree. We note that we held $n$ constant at $n = 3.2$ (see Eq. \ref{eq:power_law}), consistent with the values from \citet{suf20}, and we did not attempt to match the observed polarization degree in detail.

\subsection{\textsc{SPH} models}

Based on near-infrared (1.6 to 2.4 μm) spectrointerferomic data, Che et al. (2012) suggested that the direction of the secondary star’s orbital motion vector is opposite to that of the disk’s rotation. In such a retrograde orbit, as the secondary is travelling toward the observer, the disk is rotating away. Using near-infrared interferometry and radio mm observations, \citet{ste12} confirmed this finding. The direction of the secondary star's motion relative to the rotating disk has important implications on the expected variations in all observables. A retrograde orbit reduces the interaction times between the secondary star and individual particles in the outer regions of the disk, therefore reducing the strength of the tidal effects \citep{ove24}.

To investigate the effects of the secondary's direction on disk geometry, size, and any potential disruptions at closest approach, an \textsc{SPH} model was run for both the prograde and the retrograde cases, where the disk and the secondary's orbit were coplanar. The parameters used in our \textsc{SPH} models are shown in Table \ref{sph_params}. These values represent the most up to date stellar and orbital parameters, primarily from \citet{suf20} and \citet{tyc11}. The mass and radius of the secondary star are taken from the interpolated values for a B2 star in \citet{sil14}, based on the values given by \citet{cox00}. The mass loss rate utilized here is consistent with the rate from \citet{suf22}. The value adopted for $\alpha$ has been commonly used in previous studies using this code and falls within the lower end of estimates made by \citet{rim18} and \citet{gho18}. Our simulations followed the disk evolution for three periastron passages, beginning at apastron half a period before the first periastron. While an extensive grid of \textsc{SPH} models was beyond the scope of this study, these models aid in the interpretation of the observational data presented earlier. 

\begin{deluxetable}{cc}
\tabletypesize{\small}
\tablewidth{0pt} 
\tablecaption{Parameters of the \textsc{SPH} models. \label{sph_params}}
\tablehead{\colhead{Parameter} & \colhead{Value}}
\startdata 
Primary Mass & 15.3 M$_{\odot}$ \\
Secondary Mass & 9.11 M$_{\odot}$ \\
        Primary Radius & 6.8 R$_{\odot}$ \\
        Secondary Radius & 5.33 R$_{\odot}$ \\
        Orbital Period & 3951 days \\
        Eccentricity & 0.938 \\
        $\alpha$ & 0.1 \\  
        Mass loss rate & $1 \times
        10^{-7}~$M$_{\odot}$ yr$^{-1}$  \\
\enddata
\end{deluxetable}

\begin{figure}
\gridline{\fig{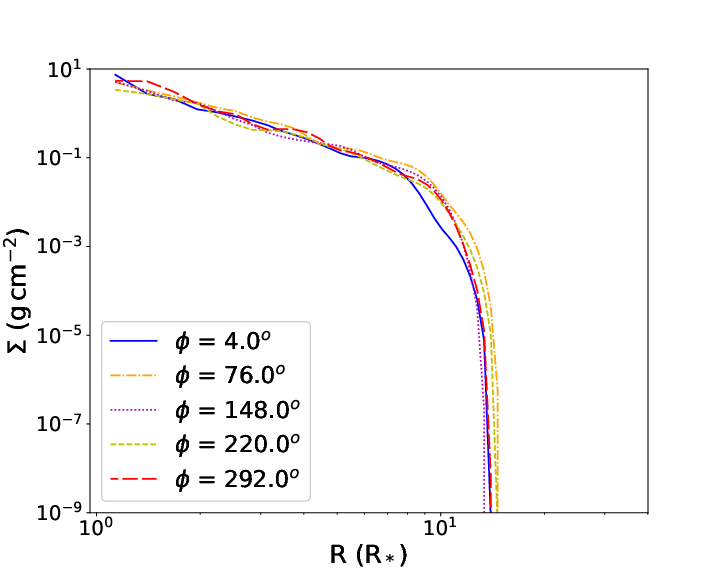}{0.4\textwidth}{(a)}}
\gridline{\fig{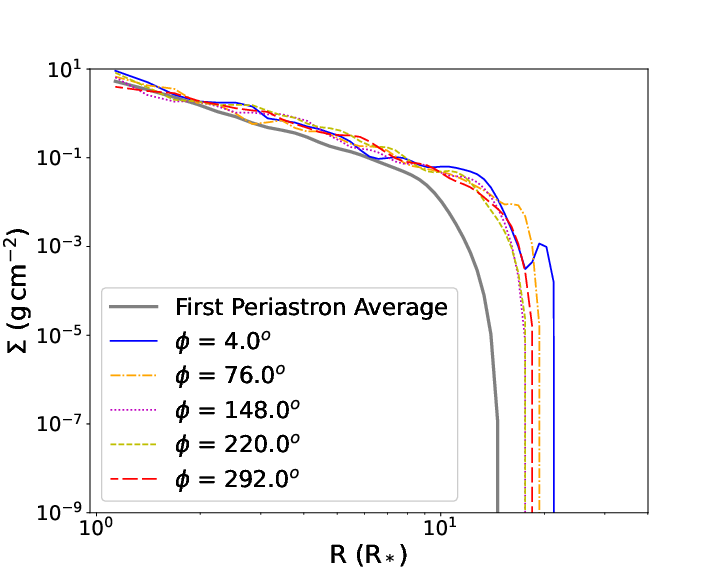}{0.4\textwidth}{(b)}}
\gridline{\fig{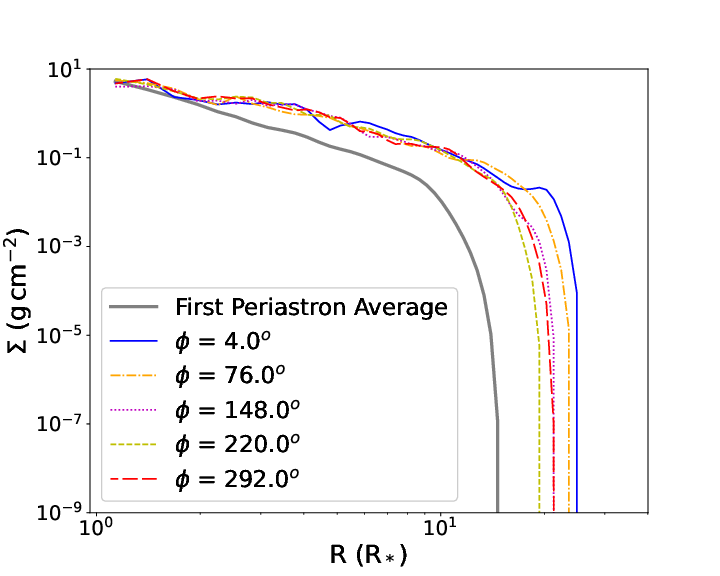}{0.4\textwidth}{(c)}}
\caption{Plots of surface density vs. radius, from the coplanar retrograde \textsc{SPH} simulation for the three periastron years using the paramters from Table \ref{sph_params}. (a) is for the first periastron passage, while (b) and (c) are for the second and third, respectively. The azimuthal angle $\phi$ is measured with respect to the binary's semi-major axis, where $\phi = 0^{\circ}$ is in the direction of periastron.}
\label{fig:sigma_rstar}
\end{figure}

The \textsc{SPH} models show changes in disk structure and spatial extent at each periastron passage. Fig. \ref{fig:sigma_rstar} plots the surface density vs. radius for various values of the azimuthal angle $\phi$ during the three periastra for the retrograde case. The angle $\phi$ is measured with respect to the binary's semi-major axis, where $\phi = 180^{\circ}$ is in the direction of apastron. Therefore, the $\phi = 0^{\circ}$ line probes the density along the direction of the periastron, when the secondary star is at its closest approach. For both the prograde and the retrograde models, the outer limit of the disk has grown between each periastron. In addition, the slope of surface density vs. radius becomes shallower at each consecutive periastron. These two facts together indicate that the disk has a larger radius at each successive periastron event, consistent with the  observational findings based on H$\alpha$ spectroscopy and polarimetry discussed above. 

Both the prograde and retrograde models also indicate that the disk becomes more asymmetric with each passing, although this effect is less pronounced in the retrograde case. In both cases, the surface density shows perturbations at periastron once the outer limit of the disk approaches $\approx$ 20 $R_{\star}$. This can be taken as an approximate value for the radius where the disk becomes perturbed by the secondary star. The values for the radius of the disk in the second and third periastra are lower than the estimated values for 2011 and 2022 discussed above; this may be because the estimate is approximate, as it does not take into consideration the diffuse emission. In the future, it would be interesting to run a more extensive set of \textsc{SPH} models with different values for the scaling parameter $\alpha$. As discussed in \citet{bro19} and \citet{pan16}, higher values of $\alpha$ allow the disk material to flow outwards more quickly, enabling faster disk growth for the same mass loss rate. However, the overall trend of disk growth and increased perturbations at each closest approach supports the observations discussed previously.

\begin{figure}[htb!]
\plotone{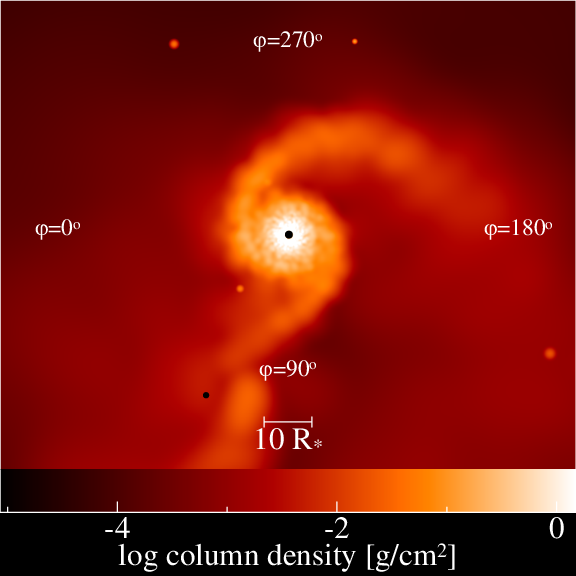}
\caption{Top-down snapshot of an \textsc{SPH} simulation based on the parameters of $\delta$ Sco shown in Table \ref{sph_params}, in which the disk is prograde with respect to the binary orbit. This image is taken immediately after the second periastron passage. The colour denotes the column density as shown in the colourbar. The primary and secondary stars are denoted by small black points, and the motion of the secondary star is counterclockwise. The brown dots represent rogue medium-density particles that have been launched from the disk.}
\label{fig:SPH_prograde}
\end{figure}

\begin{figure}[htb!]
\plotone{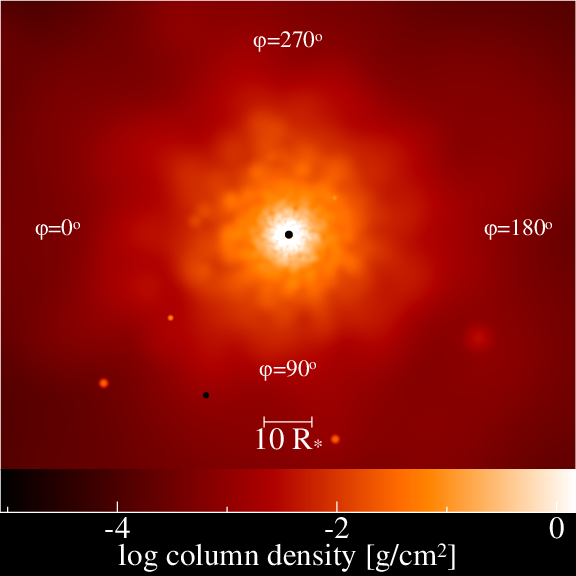}
\caption{Top-down snapshot of an \textsc{SPH} simulation based on the parameters of $\delta$ Sco shown in Table \ref{sph_params}, in which the disk is retrograde with respect to the binary orbit. This image is taken immediately after the second periastron passage. The format is the same as Fig. \ref{fig:SPH_prograde}.}
\label{fig:SPH_retrograde}
\end{figure}

Fig. \ref{fig:SPH_prograde} shows a snapshot of the prograde simulation, immediately after the second periastron passage. A two-armed ($m=2$) spiral density enhancement is clearly visible in the outer regions of the disk. This density wave is similar to those produced in \textsc{SPH} simulations of binary Be stars by \citet{cyr20}. These arms dissipate within 0.02 to 0.03 orbital periods (80 to 120 days) of closest approach, once the distance between the secondary star and the disk is sufficiently large for the tidal torques to become negligible. Fig. \ref{fig:SPH_retrograde} shows a snapshot at the same point in time as in Fig. \ref{fig:SPH_prograde}, but for the retrograde case. The secondary star does not produce a clear spiral arm structure in this case, as the interaction time between the secondary star and individual disk particles is shorter than in the prograde case due to their relative velocities. Overall, our models for the retrograde case are consistent with the observed photometric, spectral and polarimetric data near periastron for the disk parameters we used (see Table \ref{sph_params}). 

\section{Conclusions} \label{sec:conclusion}

 Using photometric, spectral, and polarimetric data to trace the behaviour of the Be binary $\delta$ Sco through its three most recent periastron events, we can create a cohesive description of the disk’s physical characteristics at each encounter with the secondary star. While we focused on the behaviour of the disk during each close encounter, the system’s high eccentricity and long period necessitated the consideration of the disk’s evolution during the intervals between each event. The disk evolved independently of the secondary star between each passage, leading to different responses to the presence of the secondary at each closest approach.
 
During the 2000 periastron, the disk was young and relatively small. It experienced truncation from the passage of the secondary star, indicated by a drop in brightness and increase in H$\alpha$ EW and FWHM around closest approach as seen in Fig. \ref{fig:periastra_obs}, but otherwise was relatively unperturbed. The polarization and $V$-band photometry shown in Fig. \ref{fig:polarization_longterm} indicate increasing disk density leading up to 2011, consistent with a building disk. This is further evidenced by the overall increase in system brightness and stronger H$\alpha$ emission shown in Fig. \ref{fig:periastra_obs}. In 2011, the H$\alpha$ line profile in Fig. \ref{fig:spectrum_position} and He\,\textsc{i} 6678 line (see Fig. \ref{fig:he_lines}) exhibit significant asymmetries near minimum separation, indicating perturbations that propagated to the inner regions of the disk, but these damped out shortly after periastron. Polarization data are consistent with disk growth in the years following 2011. By 2022, the disk had grown again in extent, as shown by continued increase in system brightness and H$\alpha$ emission in addition to a further drop in peak separation, shown in Fig. \ref{fig:periastra_obs}. Once again, the H$\alpha$ line showed significant asymmetries around closest approach, as demonstrated by the EW ratios in Fig. \ref{fig:periastra_obs} and the profiles shown in Fig. \ref{fig:spectrum_position}.

Overall, the passage of the secondary star produces increasingly larger perturbations in the disk with each passage, due to the growth of the disk between periastra. Our model with a retrograde coplanar disk, shown in Fig. \ref{fig:SPH_retrograde} indicates that these perturbations are short-lived, as the tidal torques quickly become negligible due to the large eccentricity of the system. One of the features of our data that remain unexplained is the drop in polarizaiton after 2013. In the future, it would be interesting to vary the viscosity parameter $\alpha$ in an extensive suite of \textsc{SPH} models.

\begin{acknowledgments}

The authors would like to thank the referee, Dietrich Baade, for his comments which greatly improved the quality of this paper. This work has made use of the BeSS database, operated at LESIA, Observatoire de Meudon, France: \href{http://basebe.obspm.fr}{http://basebe.obspm.fr}. Special thanks to observer Ernst Pollmann for contributions of H$\alpha$ spectra. We thank Jonathan Labadie-Bartz for providing processed spectra containing helium lines obtained from BeSS. Ritter Observatory Public Archive is supported by the National Science Foundation Program for Research and Education with Small Telescopes (PREST). We acknowledge with thanks the variable star observations from the AAVSO International Database contributed by observers worldwide and used in this research.  C.E.J. acknowledges support through the National Science and Engineering Research Council of Canada. A.C.C. acknowledges support from CNPq (grant 311446/2019-1) and FAPESP (grants 2018/04055-8 and 2019/13354-1). This work was made possible through the use of the Shared Hierarchical Academic Research Computing Network (SHARCNET). G.W.H. thanks Lou Boyd, Director of Fairborn Observatory, for his decades of support for our robotic telescopes. C.T. would like to acknowledge, with thanks, FRCE grant support from Central Michigan University. We acknowledge the use of SPLASH \citep{pri07} for rendering and visualization of our figures. 

\end{acknowledgments}

\clearpage




\bibliography{sample631}{}

\begin{thebibliography}{}
\expandafter\ifx\csname natexlab\endcsname\relax\def\natexlab#1{#1}\fi
\providecommand{\url}[1]{\href{#1}{#1}}
\providecommand{\dodoi}[1]{doi:~\href{http://doi.org/#1}{\nolinkurl{#1}}}
\providecommand{\doeprint}[1]{\href{http://ascl.net/#1}{\nolinkurl{http://ascl.net/#1}}}
\providecommand{\doarXiv}[1]{\href{https://arxiv.org/abs/#1}{\nolinkurl{https://arxiv.org/abs/#1}}}

\bibitem[{{Baade} {et~al.}(2023){Baade}, {Labadie-Bartz}, {Rivinius}, \& {Carciofi}}]{baa23}
{Baade}, D., {Labadie-Bartz}, J., {Rivinius}, T., \& {Carciofi}, A.~C. 2023, \aap, 678, A47, \dodoi{10.1051/0004-6361/202244149}

\bibitem[{{Baade} \& {Rivinius}(2020)}]{baa20}
{Baade}, D., \& {Rivinius}, T. 2020, in Stars and their Variability Observed from Space, ed. C.~{Neiner}, W.~W. {Weiss}, D.~{Baade}, R.~E. {Griffin}, C.~C. {Lovekin}, \& A.~F.~J. {Moffat}, 35--38

\bibitem[{{Baade} {et~al.}(2018){Baade}, {Rivinius}, {Pigulski}, {Panoglou}, {Carciofi}, {Handler}, {Kuschnig}, {Martayan}, {Mehner}, {Moffat}, {Pablo}, {Popowicz}, {Rucinski}, {Wade}, {Weiss}, \& {Zwintz}}]{baa18}
{Baade}, D., {Rivinius}, T., {Pigulski}, A., {et~al.} 2018, in 3rd BRITE Science Conference, ed. G.~A. {Wade}, D.~{Baade}, J.~A. {Guzik}, \& R.~{Smolec}, Vol.~8, 69--76, \dodoi{10.48550/arXiv.1708.08413}

\bibitem[{Babusiaux {et~al.}(2023)Babusiaux, Fabricius, Khanna, Muraveva, Reyl{\'{e} }, Spoto, Vallenari, Luri, Arenou, {\'{A}}lvarez, Anders, Antoja, Balbinot, Barache, Bauchet, Bossini, Busonero, Cantat-Gaudin, Carrasco, Dafonte, Diakit{\'{e}}, Figueras, Garcia-Gutierrez, Garofalo, Helmi, Jim{\'{e}}nez-Arranz, Jordi, Kervella, Kostrzewa-Rutkowska, Leclerc, Licata, Manteiga, Masip, Mongui{\'{o}}, Ramos, Robichon, Robin, Romero-G{\'{o}}mez, S{\'{a}}ez, Santove{\~{n}}a, Spina, Elipe, \& Weiler}]{bab23}
Babusiaux, C., Fabricius, C., Khanna, S., {et~al.} 2023, \aap, 674, A32, \dodoi{10.1051/0004-6361/202243790}

\bibitem[{{Baldwin}(1940)}]{bal40}
{Baldwin}, R.~B. 1940, \apj, 92, 82, \dodoi{10.1086/144203}

\bibitem[{{Bate} {et~al.}(1995){Bate}, {Bonnell}, \& {Price}}]{bat95}
{Bate}, M.~R., {Bonnell}, I.~A., \& {Price}, N.~M. 1995, \mnras, 277, 362, \dodoi{10.1093/mnras/277.2.362}

\bibitem[{{Bedding}(1993)}]{bed93}
{Bedding}, T.~R. 1993, \aj, 106, 768, \dodoi{10.1086/116684}

\bibitem[{{Benz}(1990)}]{ben90b}
{Benz}, W. 1990, in Numerical Modelling of Nonlinear Stellar Pulsations Problems and Prospects, ed. J.~R. {Buchler}, 269

\bibitem[{{Benz} {et~al.}(1990){Benz}, {Bowers}, {Cameron}, \& {Press}}]{ben90a}
{Benz}, W., {Bowers}, R.~L., {Cameron}, A.~G.~W., \& {Press}, W.~H.~. 1990, \apj, 348, 647, \dodoi{10.1086/168273}

\bibitem[{{Bernacca} \& {Perinotto}(1970)}]{ber70}
{Bernacca}, P.~L., \& {Perinotto}, M. 1970, Contributi dell'Osservatorio Astrofisica dell'Universita di Padova in Asiago, 239, 1

\bibitem[{{Bjorkman}(2012)}]{bjo12}
{Bjorkman}, J.~E. 2012, in Astronomical Society of the Pacific Conference Series, Vol. 464, Circumstellar Dynamics at High Resolution, ed. A.~C. {Carciofi} \& T.~{Rivinius}, 85

\bibitem[{{Bjorkman} {et~al.}(2002){Bjorkman}, {Miroshnichenko}, {McDavid}, \& {Pogrosheva}}]{bjo02}
{Bjorkman}, K.~S., {Miroshnichenko}, A.~S., {McDavid}, D., \& {Pogrosheva}, T.~M. 2002, \apj, 573, 812, \dodoi{10.1086/340751}

\bibitem[{{Bodensteiner} {et~al.}(2020){Bodensteiner}, {Shenar}, \& {Sana}}]{bod20}
{Bodensteiner}, J., {Shenar}, T., \& {Sana}, H. 2020, \aap, 641, A42, \dodoi{10.1051/0004-6361/202037640}

\bibitem[{{Borre} {et~al.}(2020){Borre}, {Baade}, {Pigulski}, {Panoglou}, {Weiss}, {Rivinius}, {Handler}, {Moffat}, {Popowicz}, {Wade}, {Weiss}, \& {Zwintz}}]{bor20}
{Borre}, C.~C., {Baade}, D., {Pigulski}, A., {et~al.} 2020, \aap, 635, A140, \dodoi{10.1051/0004-6361/201937062}

\bibitem[{{Brown} {et~al.}(2019){Brown}, {Coe}, {Ho}, \& {Okazaki}}]{bro19}
{Brown}, R.~O., {Coe}, M.~J., {Ho}, W.~C.~G., \& {Okazaki}, A.~T. 2019, \mnras, 488, 387, \dodoi{10.1093/mnras/stz1757}

\bibitem[{{Carciofi}(2011)}]{car11}
{Carciofi}, A.~C. 2011, in Active OB Stars: Structure, Evolution, Mass Loss, and Critical Limits, ed. C.~{Neiner}, G.~{Wade}, G.~{Meynet}, \& G.~{Peters}, Vol. 272, 325--336, \dodoi{10.1017/S1743921311010738}

\bibitem[{{Carciofi} \& {Bjorkman}(2006)}]{car06a}
{Carciofi}, A.~C., \& {Bjorkman}, J.~E. 2006, \apj, 639, 1081, \dodoi{10.1086/499483}

\bibitem[{{Carciofi} {et~al.}(2007){Carciofi}, {Magalh{\~a}es}, {Leister}, {Bjorkman}, \& {Levenhagen}}]{car07}
{Carciofi}, A.~C., {Magalh{\~a}es}, A.~M., {Leister}, N.~V., {Bjorkman}, J.~E., \& {Levenhagen}, R.~S. 2007, \apjl, 671, L49, \dodoi{10.1086/524772}

\bibitem[{{Carciofi} {et~al.}(2009){Carciofi}, {Okazaki}, {Le Bouquin}, {{\v{S}}tefl}, {Rivinius}, {Baade}, {Bjorkman}, \& {Hummel}}]{car09}
{Carciofi}, A.~C., {Okazaki}, A.~T., {Le Bouquin}, J.~B., {et~al.} 2009, \aap, 504, 915, \dodoi{10.1051/0004-6361/200810962}

\bibitem[{{Carciofi} {et~al.}(2006){Carciofi}, {Miroshnichenko}, {Kusakin}, {Bjorkman}, {Bjorkman}, {Marang}, {Kuratov}, {Garc{\'\i}a-Lario}, {Calder{\'o}n}, {Fabregat}, \& {Magalh{\~a}es}}]{car06b}
{Carciofi}, A.~C., {Miroshnichenko}, A.~S., {Kusakin}, A.~V., {et~al.} 2006, \apj, 652, 1617, \dodoi{10.1086/507935}

\bibitem[{{Che} {et~al.}(2012){Che}, {Monnier}, {Tycner}, {Kraus}, {Zavala}, {Baron}, {Pedretti}, {ten Brummelaar}, {McAlister}, {Ridgway}, {Sturmann}, {Sturmann}, \& {Turner}}]{che12}
{Che}, X., {Monnier}, J.~D., {Tycner}, C., {et~al.} 2012, \apj, 757, 29, \dodoi{10.1088/0004-637X/757/1/29}

\bibitem[{{Collins}(1987)}]{col87}
{Collins}, George~W., I. 1987, in IAU Colloq. 92: Physics of Be Stars, ed. A.~{Slettebak} \& T.~P. {Snow}, 3

\bibitem[{{Cote} \& {van Kerkwijk}(1993)}]{cot93}
{Cote}, J., \& {van Kerkwijk}, M.~H. 1993, \aap, 274, 870

\bibitem[{{Cotton} {et~al.}(2019){Cotton}, {Marshall}, {Frisch}, {Kedziora-Chudzer}, {Bailey}, {Bott}, {Wright}, {Wyatt}, \& {Kennedy}}]{cot19}
{Cotton}, D.~V., {Marshall}, J.~P., {Frisch}, P.~C., {et~al.} 2019, \mnras, 483, 3636, \dodoi{10.1093/mnras/sty3318}

\bibitem[{{Cox}(2000)}]{cox00}
{Cox}, A.~N. 2000, {Allen's astrophysical quantities} ({New York: AIP Press; Springer})

\bibitem[{{Cyr} {et~al.}(2020){Cyr}, {Jones}, {Carciofi}, {Steckel}, {Tycner}, \& {Okazaki}}]{cyr20}
{Cyr}, I.~H., {Jones}, C.~E., {Carciofi}, A.~C., {et~al.} 2020, \mnras, 497, 3525, \dodoi{10.1093/mnras/staa2176}

\bibitem[{{Cyr} {et~al.}(2017){Cyr}, {Jones}, {Panoglou}, {Carciofi}, \& {Okazaki}}]{cyr17}
{Cyr}, I.~H., {Jones}, C.~E., {Panoglou}, D., {Carciofi}, A.~C., \& {Okazaki}, A.~T. 2017, \mnras, 471, 596, \dodoi{10.1093/mnras/stx1427}

\bibitem[{{de Almeida} {et~al.}(2020){de Almeida}, {Meilland}, {Domiciano de Souza}, {Stee}, {Mourard}, {Nardetto}, {Ligi}, {Tallon-Bosc}, {Faes}, {Carciofi}, {Bednarski}, {Mota}, {Turner}, \& {ten Brummelaar}}]{dea20}
{de Almeida}, E.~S.~G., {Meilland}, A., {Domiciano de Souza}, A., {et~al.} 2020, \aap, 636, A110, \dodoi{10.1051/0004-6361/201936039}

\bibitem[{{de Wit} {et~al.}(2006){de Wit}, {Lamers}, {Marquette}, \& {Beaulieu}}]{dew06}
{de Wit}, W.~J., {Lamers}, H.~J.~G.~L.~M., {Marquette}, J.~B., \& {Beaulieu}, J.~P. 2006, \aap, 456, 1027, \dodoi{10.1051/0004-6361:20065137}

\bibitem[{{Dodd} {et~al.}(2024){Dodd}, {Oudmaijer}, {Radley}, {Vioque}, \& {Frost}}]{dod24}
{Dodd}, J.~M., {Oudmaijer}, R.~D., {Radley}, I.~C., {Vioque}, M., \& {Frost}, A.~J. 2024, \mnras, 527, 3076, \dodoi{10.1093/mnras/stad3105}

\bibitem[{{Draper} {et~al.}(2014){Draper}, {Wisniewski}, {Bjorkman}, {Meade}, {Haubois}, {Mota}, {Carciofi}, \& {Bjorkman}}]{dra14}
{Draper}, Z.~H., {Wisniewski}, J.~P., {Bjorkman}, K.~S., {et~al.} 2014, \apj, 786, 120, \dodoi{10.1088/0004-637X/786/2/120}

\bibitem[{{Eggleton}(1983)}]{egg83}
{Eggleton}, P.~P. 1983, \apj, 268, 368, \dodoi{10.1086/160960}

\bibitem[{{Fr{\'e}mat} {et~al.}(2005){Fr{\'e}mat}, {Zorec}, {Hubert}, \& {Floquet}}]{fre05}
{Fr{\'e}mat}, Y., {Zorec}, J., {Hubert}, A.~M., \& {Floquet}, M. 2005, \aap, 440, 305, \dodoi{10.1051/0004-6361:20042229}

\bibitem[{{Ghoreyshi} {et~al.}(2021){Ghoreyshi}, {Carciofi}, {Jones}, {Faes}, {Baade}, \& {Rivinius}}]{gho21}
{Ghoreyshi}, M.~R., {Carciofi}, A.~C., {Jones}, C.~E., {et~al.} 2021, \apj, 909, 149, \dodoi{10.3847/1538-4357/abdd1e}

\bibitem[{{Ghoreyshi} {et~al.}(2018){Ghoreyshi}, {Carciofi}, {R{\'\i}mulo}, {Vieira}, {Faes}, {Baade}, {Bjorkman}, {Otero}, \& {Rivinius}}]{gho18}
{Ghoreyshi}, M.~R., {Carciofi}, A.~C., {R{\'\i}mulo}, L.~R., {et~al.} 2018, \mnras, 479, 2214, \dodoi{10.1093/mnras/sty1577}

\bibitem[{{Hall}(1958)}]{hal58}
{Hall}, J.~S. 1958, Publications of the U.S. Naval Observatory Second Series, 17, 275

\bibitem[{{Halonen} \& {Jones}(2013)}]{hal13}
{Halonen}, R.~J., \& {Jones}, C.~E. 2013, \apj, 765, 17, \dodoi{10.1088/0004-637X/765/1/17}

\bibitem[{{Hanuschik}(1989)}]{han89}
{Hanuschik}, R.~W. 1989, \apss, 161, 61, \dodoi{10.1007/BF00653238}

\bibitem[{{Haubois} {et~al.}(2012){Haubois}, {Carciofi}, {Rivinius}, {Okazaki}, \& {Bjorkman}}]{hau12}
{Haubois}, X., {Carciofi}, A.~C., {Rivinius}, T., {Okazaki}, A.~T., \& {Bjorkman}, J.~E. 2012, \apj, 756, 156, \dodoi{10.1088/0004-637X/756/2/156}

\bibitem[{{Haubois} {et~al.}(2014){Haubois}, {Mota}, {Carciofi}, {Draper}, {Wisniewski}, {Bednarski}, \& {Rivinius}}]{hau14}
{Haubois}, X., {Mota}, B.~C., {Carciofi}, A.~C., {et~al.} 2014, \apj, 785, 12, \dodoi{10.1088/0004-637X/785/1/12}

\bibitem[{{Hayes} \& {Guinan}(1984)}]{gui84}
{Hayes}, D.~P., \& {Guinan}, E.~F. 1984, \apj, 279, 721, \dodoi{10.1086/161938}

\bibitem[{{Henry}(1999)}]{hen99}
{Henry}, G.~W. 1999, \pasp, 111, 845, \dodoi{10.1086/316388}

\bibitem[{{Huang}(1972)}]{hua72}
{Huang}, S.-S. 1972, \apj, 171, 549, \dodoi{10.1086/151309}

\bibitem[{{Hummel}(1994)}]{hum94}
{Hummel}, W. 1994, \aap, 289, 458

\bibitem[{{Hummel} \& {Vrancken}(1995)}]{hum95}
{Hummel}, W., \& {Vrancken}, M. 1995, \aap, 302, 751

\bibitem[{{Hutter} {et~al.}(2021){Hutter}, {Tycner}, {Zavala}, {Benson}, {Hummel}, \& {Zirm}}]{hut21}
{Hutter}, D.~J., {Tycner}, C., {Zavala}, R.~T., {et~al.} 2021, \apjs, 257, 69, \dodoi{10.3847/1538-4365/ac23cb}

\bibitem[{{Jaschek} {et~al.}(1981){Jaschek}, {Slettebak}, \& {Jaschek}}]{jas81}
{Jaschek}, M., {Slettebak}, A., \& {Jaschek}, C. 1981, Be Star Newsletter, 4, 9

\bibitem[{{Jones} {et~al.}(2011){Jones}, {Tycner}, \& {Smith}}]{jon11}
{Jones}, C.~E., {Tycner}, C., \& {Smith}, A.~D. 2011, \aj, 141, 150, \dodoi{10.1088/0004-6256/141/5/150}

\bibitem[{{Klement} {et~al.}(2019){Klement}, {Carciofi}, {Rivinius}, {Ignace}, {Matthews}, {Torstensson}, {Gies}, {Vieira}, {Richardson}, {Domiciano de Souza}, {Bjorkman}, {Hallinan}, {Faes}, {Mota}, {Gullingsrud}, {de Breuck}, {Kervella}, {Cur{\'e}}, \& {Gunawan}}]{kle19}
{Klement}, R., {Carciofi}, A.~C., {Rivinius}, T., {et~al.} 2019, \apj, 885, 147, \dodoi{10.3847/1538-4357/ab48e7}

\bibitem[{{Kopal}(1959)}]{kop59}
{Kopal}, Z. 1959, {Close binary systems} ({The International Astrophysics Series, London: Chapman \& Hall})

\bibitem[{{Labadie-Bartz} {et~al.}(2022){Labadie-Bartz}, {Carciofi}, {Henrique de Amorim}, {Rubio}, {Luiz Figueiredo}, {Ticiani dos Santos}, \& {Thomson-Paressant}}]{lab22}
{Labadie-Bartz}, J., {Carciofi}, A.~C., {Henrique de Amorim}, T., {et~al.} 2022, \aj, 163, 226, \dodoi{10.3847/1538-3881/ac5abd}

\bibitem[{{Lee} {et~al.}(1991){Lee}, {Osaki}, \& {Saio}}]{lee91}
{Lee}, U., {Osaki}, Y., \& {Saio}, H. 1991, \mnras, 250, 432, \dodoi{10.1093/mnras/250.2.432}

\bibitem[{{Magalhaes} {et~al.}(1984){Magalhaes}, {Benedetti}, \& {Roland}}]{mag84}
{Magalhaes}, A.~M., {Benedetti}, E., \& {Roland}, E.~H. 1984, \pasp, 96, 383, \dodoi{10.1086/131351}

\bibitem[{{Magalhaes} {et~al.}(1996){Magalhaes}, {Rodrigues}, {Margoniner}, {Pereyra}, \& {Heathcote}}]{mag96}
{Magalhaes}, A.~M., {Rodrigues}, C.~V., {Margoniner}, V.~E., {Pereyra}, A., \& {Heathcote}, S. 1996, in Astronomical Society of the Pacific Conference Series, Vol.~97, Polarimetry of the Interstellar Medium, ed. W.~G. {Roberge} \& D.~C.~B. {Whittet}, 118

\bibitem[{{Marchant} \& {Bodensteiner}(2023)}]{mar23}
{Marchant}, P., \& {Bodensteiner}, J. 2023, arXiv e-prints, arXiv:2311.01865, \dodoi{10.48550/arXiv.2311.01865}

\bibitem[{{Marr} {et~al.}(2021){Marr}, {Jones}, {Carciofi}, {Rubio}, {Mota}, {Ghoreyshi}, {Hatfield}, \& {R{\'\i}mulo}}]{mar21}
{Marr}, K.~C., {Jones}, C.~E., {Carciofi}, A.~C., {et~al.} 2021, \apj, 912, 76, \dodoi{10.3847/1538-4357/abed4c}

\bibitem[{{Marr} {et~al.}(2022){Marr}, {Jones}, {Tycner}, {Carciofi}, \& {Silva}}]{mar22}
{Marr}, K.~C., {Jones}, C.~E., {Tycner}, C., {Carciofi}, A.~C., \& {Silva}, A.~C.~F. 2022, \apj, 928, 145, \dodoi{10.3847/1538-4357/ac551b}

\bibitem[{{Martin} {et~al.}(1992){Martin}, {Adamson}, {Whittet}, {Hough}, {Bailey}, {Kim}, {Sato}, {Tamura}, \& {Yamashita}}]{mar92}
{Martin}, P.~G., {Adamson}, A.~J., {Whittet}, D. C.~B., {et~al.} 1992, \apj, 392, 691

\bibitem[{{Martin} {et~al.}(2011){Martin}, {Pringle}, {Tout}, \& {Lubow}}]{mar11}
{Martin}, R.~G., {Pringle}, J.~E., {Tout}, C.~A., \& {Lubow}, S.~H. 2011, \mnras, 416, 2827, \dodoi{10.1111/j.1365-2966.2011.19231.x}

\bibitem[{{Meilland} {et~al.}(2013){Meilland}, {Stee}, {Spang}, {Malbet}, {Massi}, \& {Schertl}}]{mei13}
{Meilland}, A., {Stee}, P., {Spang}, A., {et~al.} 2013, \aap, 550, L5, \dodoi{10.1051/0004-6361/201220712}

\bibitem[{{Meilland} {et~al.}(2007){Meilland}, {Stee}, {Vannier}, {Millour}, {Domiciano de Souza}, {Malbet}, {Martayan}, {Paresce}, {Petrov}, {Richichi}, \& {Spang}}]{mei07}
{Meilland}, A., {Stee}, P., {Vannier}, M., {et~al.} 2007, \aap, 464, 59, \dodoi{10.1051/0004-6361:20064848}

\bibitem[{{Meilland} {et~al.}(2011){Meilland}, {Delaa}, {Stee}, {Kanaan}, {Millour}, {Mourard}, {Bonneau}, {Petrov}, {Nardetto}, {Marcotto}, {Roussel}, {Clausse}, {Perraut}, {McAlister}, {ten Brummelaar}, {Sturmann}, {Sturmann}, {Turner}, {Ridgway}, {Farrington}, \& {Goldfinger}}]{mei11}
{Meilland}, A., {Delaa}, O., {Stee}, P., {et~al.} 2011, \aap, 532, A80, \dodoi{10.1051/0004-6361/201116798}

\bibitem[{{Millan-Gabet} {et~al.}(2010){Millan-Gabet}, {Monnier}, {Touhami}, {Gies}, {Hesselbach}, {Pedretti}, {Thureau}, {Zhao}, {ten Brummelaar}, \& {CHARA Group}}]{mil10}
{Millan-Gabet}, R., {Monnier}, J.~D., {Touhami}, Y., {et~al.} 2010, \apj, 723, 544, \dodoi{10.1088/0004-637X/723/1/544}

\bibitem[{{Miroshnichenko}(2016)}]{mir16}
{Miroshnichenko}, A.~S. 2016, in Astronomical Society of the Pacific Conference Series, Vol. 506, Bright Emissaries: Be Stars as Messengers of Star-Disk Physics, ed. T.~A.~A. {Sigut} \& C.~E. {Jones}, 71

\bibitem[{{Miroshnichenko} {et~al.}(2001){Miroshnichenko}, {Fabregat}, {Bjorkman}, {Knauth}, {Morrison}, {Tarasov}, {Reig}, {Negueruela}, \& {Blay}}]{mir01}
{Miroshnichenko}, A.~S., {Fabregat}, J., {Bjorkman}, K.~S., {et~al.} 2001, \aap, 377, 485, \dodoi{10.1051/0004-6361:20010911}

\bibitem[{{Miroshnichenko} {et~al.}(2003){Miroshnichenko}, {Bjorkman}, {Morrison}, {Wisniewski}, {Manset}, {Levato}, {Grosso}, {Pollmann}, {Buil}, \& {Knauth}}]{mir03}
{Miroshnichenko}, A.~S., {Bjorkman}, K.~S., {Morrison}, N.~D., {et~al.} 2003, \aap, 408, 305, \dodoi{10.1051/0004-6361:20030965}

\bibitem[{{Miroshnichenko} {et~al.}(2013){Miroshnichenko}, {Pasechnik}, {Manset}, {Carciofi}, {Rivinius}, {{\v{S}}tefl}, {Gvaramadze}, {Ribeiro}, {Fernando}, {Garrel}, {Knapen}, {Buil}, {Heathcote}, {Pollmann}, {Mauclaire}, {Thizy}, {Martin}, {Zharikov}, {Okazaki}, {Gandet}, {Eversberg}, \& {Reinecke}}]{mir13}
{Miroshnichenko}, A.~S., {Pasechnik}, A.~V., {Manset}, N., {et~al.} 2013, \apj, 766, 119, \dodoi{10.1088/0004-637X/766/2/119}

\bibitem[{{Miroshnichenko} {et~al.}(2023){Miroshnichenko}, {Chari}, {Danford}, {Prendergast}, {Aarnio}, {Andronov}, {Chinarova}, {Lytle}, {Amantayeva}, {Gabitova}, {Vaidman}, {Baktybayev}, \& {Khokhlov}}]{mir23}
{Miroshnichenko}, A.~S., {Chari}, R., {Danford}, S., {et~al.} 2023, Galaxies, 11, 83, \dodoi{10.3390/galaxies11040083}

\bibitem[{{Neiner} {et~al.}(2011){Neiner}, {de Batz}, {Cochard}, {Floquet}, {Mekkas}, \& {Desnoux}}]{nei11}
{Neiner}, C., {de Batz}, B., {Cochard}, F., {et~al.} 2011, \aj, 142, 149, \dodoi{10.1088/0004-6256/142/5/149}

\bibitem[{{Nordsieck} \& {Harris}(1996)}]{nord96}
{Nordsieck}, K.~H., \& {Harris}, W. 1996, in Astronomical Society of the Pacific Conference Series, Vol.~97, Polarimetry of the Interstellar Medium, ed. W.~G. {Roberge} \& D.~C.~B. {Whittet}, 100

\bibitem[{{Okazaki}(1997)}]{oka97}
{Okazaki}, A.~T. 1997, \aap, 318, 548

\bibitem[{{Okazaki}(2007)}]{oka07}
{Okazaki}, A.~T. 2007, in Astronomical Society of the Pacific Conference Series, Vol. 367, Massive Stars in Interactive Binaries, ed. N.~{St. -Louis} \& A.~F.~J. {Moffat}, 485

\bibitem[{Okazaki {et~al.}(2002)Okazaki, Bate, Ogilvie, \& Pringle}]{oka02}
Okazaki, A.~T., Bate, M.~R., Ogilvie, G.~I., \& Pringle, J.~E. 2002, \mnras, 337, 967, \dodoi{10.1046/j.1365-8711.2002.05960.x}

\bibitem[{{Overton} {et~al.}(2024){Overton}, {Martin}, {Lubow}, \& {Lepp}}]{ove24}
{Overton}, M., {Martin}, R.~G., {Lubow}, S.~H., \& {Lepp}, S. 2024, \mnras, 528, L106, \dodoi{10.1093/mnrasl/slad172}

\bibitem[{{Panoglou} {et~al.}(2016){Panoglou}, {Carciofi}, {Vieira}, {Cyr}, {Jones}, {Okazaki}, \& {Rivinius}}]{pan16}
{Panoglou}, D., {Carciofi}, A.~C., {Vieira}, R.~G., {et~al.} 2016, \mnras, 461, 2616, \dodoi{10.1093/mnras/stw1508}

\bibitem[{{Panoglou} {et~al.}(2018){Panoglou}, {Faes}, {Carciofi}, {Okazaki}, {Baade}, {Rivinius}, \& {Borges Fernandes}}]{pan18}
{Panoglou}, D., {Faes}, D.~M., {Carciofi}, A.~C., {et~al.} 2018, \mnras, 473, 3039, \dodoi{10.1093/mnras/stx2497}

\bibitem[{{Peters} {et~al.}(2013){Peters}, {Pewett}, {Gies}, {Touhami}, \& {Grundstrom}}]{pet13}
{Peters}, G.~J., {Pewett}, T.~D., {Gies}, D.~R., {Touhami}, Y.~N., \& {Grundstrom}, E.~D. 2013, \apj, 765, 2, \dodoi{10.1088/0004-637X/765/1/2}

\bibitem[{{Porter} \& {Rivinius}(2003)}]{por03}
{Porter}, J.~M., \& {Rivinius}, T. 2003, \pasp, 115, 1153, \dodoi{10.1086/378307}

\bibitem[{{Price}(2007)}]{pri07}
{Price}, D.~J. 2007, \pasa, 24, 159, \dodoi{10.1071/AS07022}

\bibitem[{{Richardson} {et~al.}(2021){Richardson}, {Thizy}, {Bjorkman}, {Carciofi}, {Rubio}, {Thomas}, {Bjorkman}, {Labadie-Bartz}, {Genaro}, {Wisniewski}, {Wang}, {Gies}, {Chojnowski}, {Daly}, {Edwards}, {Fowler}, {Gullingsrud}, {Habel}, {James}, {Kehoe}, {Kuchta}, {Lane}, {Miroshnichenko}, {Mishra}, {Pablo}, {Peploski}, {Pepper}, {Rodriguez}, {Siverd}, {Stassun}, {Stevens}, {Trucks}, {Windsor}, {Wood}, {Bertrand}, {Broussat}, {Bryssinck}, {Buil}, {Charbonnel}, {de Bruin}, {Daglen}, {Desnoux}, {Dull}, {Garde}, {Graham}, {Gurney}, {Halsey}, {Fosanelli}, {Guarro Fl{\'o}}, {Houpert}, {James}, {Kreider}, {Leadbeater}, {Lester}, {Li}, {Maetz}, {Stiewing}, {Somogyi}, {Terry}, {Ubaud}, \& {Waldschlaeger}}]{ric21}
{Richardson}, N.~D., {Thizy}, O., {Bjorkman}, J.~E., {et~al.} 2021, \mnras, 508, 2002, \dodoi{10.1093/mnras/stab2759}

\bibitem[{{R{\'\i}mulo} {et~al.}(2018){R{\'\i}mulo}, {Carciofi}, {Vieira}, {Rivinius}, {Faes}, {Figueiredo}, {Bjorkman}, {Georgy}, {Ghoreyshi}, \& {Soszy{\'n}ski}}]{rim18}
{R{\'\i}mulo}, L.~R., {Carciofi}, A.~C., {Vieira}, R.~G., {et~al.} 2018, \mnras, 476, 3555, \dodoi{10.1093/mnras/sty431}

\bibitem[{{Rivinius} {et~al.}(2013){Rivinius}, {Carciofi}, \& {Martayan}}]{riv13}
{Rivinius}, T., {Carciofi}, A.~C., \& {Martayan}, C. 2013, \aapr, 21, 69, \dodoi{10.1007/s00159-013-0069-0}

\bibitem[{{Rubio} {et~al.}(2023){Rubio}, {Carciofi}, {Ticiani}, {Mota}, {Vieira}, {Faes}, {Genaro}, {de Amorim}, {Klement}, {Araya}, {Arcos}, {Cur{\'e}}, {Domiciano de Souza}, {Georgy}, {Jones}, {Suffak}, \& {Silva}}]{rub23}
{Rubio}, A.~C., {Carciofi}, A.~C., {Ticiani}, P., {et~al.} 2023, \mnras, 526, 3007, \dodoi{10.1093/mnras/stad2652}

\bibitem[{{Ru{\v{z}}djak} {et~al.}(2009){Ru{\v{z}}djak}, {Bo{\v{z}}i{\'c}}, {Harmanec}, {Fi{\v{r}}t}, {Chadima}, {Bjorkman}, {Gies}, {Kaye}, {Koubsk{\'y}}, {McDavid}, {Richardson}, {Sudar}, {{\v{S}}lechta}, {Wolf}, \& {Yang}}]{ruz09}
{Ru{\v{z}}djak}, D., {Bo{\v{z}}i{\'c}}, H., {Harmanec}, P., {et~al.} 2009, \aap, 506, 1319, \dodoi{10.1051/0004-6361/200810526}

\bibitem[{{Serkowski} {et~al.}(1975){Serkowski}, {Mathewson}, \& {Ford}}]{ser75}
{Serkowski}, K., {Mathewson}, D.~S., \& {Ford}, V.~L. 1975, \apj, 196, 261, \dodoi{10.1086/153410}

\bibitem[{{Shakura} \& {Sunyaev}(1973)}]{sha73}
{Shakura}, N.~I., \& {Sunyaev}, R.~A. 1973, \aap, 24, 337

\bibitem[{{Sigut} \& {Ghafourian}(2023)}]{sig23}
{Sigut}, T.~A.~A., \& {Ghafourian}, N.~R. 2023, \apj, 948, 34, \dodoi{10.3847/1538-4357/ac940c}

\bibitem[{{Sigut} {et~al.}(2020){Sigut}, {Mahjour}, \& {Tycner}}]{sig20}
{Sigut}, T.~A.~A., {Mahjour}, A.~K., \& {Tycner}, C. 2020, \apj, 894, 18, \dodoi{10.3847/1538-4357/ab8386}

\bibitem[{{Silaj} {et~al.}(2014){Silaj}, {Jones}, {Sigut}, \& {Tycner}}]{sil14}
{Silaj}, J., {Jones}, C.~E., {Sigut}, T.~A.~A., \& {Tycner}, C. 2014, \apj, 795, 82, \dodoi{10.1088/0004-637X/795/1/82}

\bibitem[{{Sonneborn} {et~al.}(1988){Sonneborn}, {Grady}, {Wu}, {Hayes}, {Guinan}, {Barker}, \& {Henrichs}}]{son88}
{Sonneborn}, G., {Grady}, C.~A., {Wu}, C.-C., {et~al.} 1988, \apj, 325, 784, \dodoi{10.1086/166049}

\bibitem[{{Suffak} {et~al.}(2022){Suffak}, {Jones}, \& {Carciofi}}]{suf22}
{Suffak}, M., {Jones}, C.~E., \& {Carciofi}, A.~C. 2022, \mnras, 509, 931, \dodoi{10.1093/mnras/stab3024}

\bibitem[{{Suffak} {et~al.}(2023){Suffak}, {Jones}, {Carciofi}, \& {de Amorim}}]{suf23}
{Suffak}, M.~W., {Jones}, C.~E., {Carciofi}, A.~C., \& {de Amorim}, T.~H. 2023, \mnras, 526, 782, \dodoi{10.1093/mnras/stad2781}

\bibitem[{{Suffak} {et~al.}(2020){Suffak}, {Jones}, {Tycner}, {Henry}, {Carciofi}, {Mota}, \& {Rubio}}]{suf20}
{Suffak}, M.~W., {Jones}, C.~E., {Tycner}, C., {et~al.} 2020, \apj, 890, 86, \dodoi{10.3847/1538-4357/ab68dc}

\bibitem[{Tango {et~al.}(2009)Tango, Davis, Jacob, Mendez, North, O'Byrne, Seneta, \& Tuthill}]{tan09}
Tango, W.~J., Davis, J., Jacob, A.~P., {et~al.} 2009, \mnras, 396, 842, \dodoi{10.1111/j.1365-2966.2009.14272.x}

\bibitem[{{Townsend} {et~al.}(2004){Townsend}, {Owocki}, \& {Howarth}}]{tow04}
{Townsend}, R.~H.~D., {Owocki}, S.~P., \& {Howarth}, I.~D. 2004, \mnras, 350, 189, \dodoi{10.1111/j.1365-2966.2004.07627.x}

\bibitem[{{Tycner} {et~al.}(2011){Tycner}, {Ames}, {Zavala}, {Hummel}, {Benson}, \& {Hutter}}]{tyc11}
{Tycner}, C., {Ames}, A., {Zavala}, R.~T., {et~al.} 2011, \apjl, 729, L5, \dodoi{10.1088/2041-8205/729/1/L5}

\bibitem[{{Tycner} {et~al.}(2005){Tycner}, {Lester}, {Hajian}, {Armstrong}, {Benson}, {Gilbreath}, {Hutter}, {Pauls}, \& {White}}]{tyc05}
{Tycner}, C., {Lester}, J.~B., {Hajian}, A.~R., {et~al.} 2005, \apj, 624, 359, \dodoi{10.1086/429126}

\bibitem[{{Vakili} {et~al.}(1998){Vakili}, {Mourard}, {Stee}, {Bonneau}, {Berio}, {Chesneau}, {Thureau}, {Morand}, {Labeyrie}, \& {Tallon-Bosc}}]{vak98}
{Vakili}, F., {Mourard}, D., {Stee}, P., {et~al.} 1998, \aap, 335, 261

\bibitem[{{van Leeuwen}(2007)}]{van07}
{van Leeuwen}, F. 2007, \aap, 474, 653, \dodoi{10.1051/0004-6361:20078357}

\bibitem[{{Vieira} \& {Carciofi}(2017)}]{vie17}
{Vieira}, R.~G., \& {Carciofi}, A.~C. 2017, in Astronomical Society of the Pacific Conference Series, Vol. 508, The B[e] Phenomenon: Forty Years of Studies, ed. A.~{Miroshnichenko}, S.~{Zharikov}, D.~{Kor{\v{c}}{\'a}kov{\'a}}, \& M.~{Wolf}, 33, \dodoi{10.48550/arXiv.1707.03105}

\bibitem[{{Vieira} {et~al.}(2015){Vieira}, {Carciofi}, \& {Bjorkman}}]{vie15}
{Vieira}, R.~G., {Carciofi}, A.~C., \& {Bjorkman}, J.~E. 2015, \mnras, 454, 2107, \dodoi{10.1093/mnras/stv2074}

\bibitem[{{{\v{S}}tefl} {et~al.}(2009){{\v{S}}tefl}, {Rivinius}, {Carciofi}, {Le Bouquin}, {Baade}, {Bjorkman}, {Hesselbach}, {Hummel}, {Okazaki}, {Pollmann}, {Rantakyr{\"o}}, \& {Wisniewski}}]{ste09}
{{\v{S}}tefl}, S., {Rivinius}, T., {Carciofi}, A.~C., {et~al.} 2009, \aap, 504, 929, \dodoi{10.1051/0004-6361/200811573}

\bibitem[{{{\v{S}}tefl} {et~al.}(2012){{\v{S}}tefl}, {LeBouquin}, {Rivinius}, {Baade}, {Carciofi}, {Haubois}, {Corder}, {Cur{\'e}}, \& {Kanaan}}]{ste12}
{{\v{S}}tefl}, S., {LeBouquin}, J.~B., {Rivinius}, T., {et~al.} 2012, in Astronomical Society of the Pacific Conference Series, Vol. 464, Circumstellar Dynamics at High Resolution, ed. A.~C. {Carciofi} \& T.~{Rivinius}, 197

\bibitem[{{Whittet} {et~al.}(1992){Whittet}, {Martin}, {Hough}, {Rouse}, {Bailey}, \& {Axon}}]{whi92}
{Whittet}, D. C.~B., {Martin}, P.~G., {Hough}, J.~H., {et~al.} 1992, \apj, 386, 562

\bibitem[{{Wilking} {et~al.}(1980){Wilking}, {Lebofsky}, {Martin}, {Rieke}, \& {Kemp}}]{wil80}
{Wilking}, B.~A., {Lebofsky}, M.~J., {Martin}, P.~G., {Rieke}, G.~H., \& {Kemp}, J.~C. 1980, \apj, 235, 905

\bibitem[{{Wisniewski} {et~al.}(2010){Wisniewski}, {Draper}, {Bjorkman}, {Meade}, {Bjorkman}, \& {Kowalski}}]{wis10}
{Wisniewski}, J.~P., {Draper}, Z.~H., {Bjorkman}, K.~S., {et~al.} 2010, \apj, 709, 1306, \dodoi{10.1088/0004-637X/709/2/1306}

\bibitem[{{Wood} {et~al.}(1997){Wood}, {Bjorkman}, \& {Bjorkman}}]{woo97}
{Wood}, K., {Bjorkman}, K.~S., \& {Bjorkman}, J.~E. 1997, \apj, 477, 926, \dodoi{10.1086/303747}

\bibitem[{{Zharikov} {et~al.}(2013){Zharikov}, {Miroshnichenko}, {Pollmann}, {Danford}, {Bjorkman}, {Morrison}, {Favaro}, {Guarro Fl{\'o}}, {Terry}, {Desnoux}, {Garrel}, {Martineau}, {Buchet}, {Ubaud}, {Mauclaire}, {Kalbermatten}, {Buil}, {Sawicki}, {Blank}, \& {Garde}}]{zha13}
{Zharikov}, S.~V., {Miroshnichenko}, A.~S., {Pollmann}, E., {et~al.} 2013, \aap, 560, A30, \dodoi{10.1051/0004-6361/201322114}

\end{thebibliography}
\bibliographystyle{aasjournal}



\end{document}